\documentclass{imsart}

\RequirePackage{amsthm,amsmath,amsfonts,amssymb}
\RequirePackage[authoryear]{natbib}

\RequirePackage[colorlinks,citecolor=blue,urlcolor=blue]{hyperref}
\RequirePackage{graphicx}

\startlocaldefs

\usepackage{algorithm}
\usepackage{algpseudocode}
\usepackage{hyperref}
\usepackage{url}

\renewcommand{\eqref}[1]{Equation~(\ref{#1})}
\usepackage{xparse}  
\NewDocumentCommand{\refwithlabel}{O{Equation} m}{#1~(\ref{#2})}

\theoremstyle{plain}

\newtheorem{theorem}{Theorem}[section]

\newtheorem{proposition}{Proposition}

\theoremstyle{definition}

\newcommand{\transpose}{{\mathrm{T}}}
\newcommand{\expectation}{{\mathrm{E}}}
\newcommand{\Var}{{\mathrm{var}}}
\newcommand{\Cov}{{\mathrm{cov}}}

\newcommand{\cor}{{\mathrm{cor}}}

\endlocaldefs

\begin{document}

\begin{frontmatter}
\title{Robust inference under Benford's law}
\runtitle{Robust inference under Benford's law}

\begin{aug}

\author[A]{\fnms{Lucio}~\snm{Barabesi}\ead[label=e1]{lucio.barabesi@unisi.it}\orcid{0000-0002-7210-4541}},
\author[B]{\fnms{Andrea}~\snm{Cerioli}\ead[label=e2]{andrea.cerioli@unipr.it}\orcid{0000-0002-2485-5674}},
\author[C]{\fnms{Andrea}~\snm{Cerasa}\ead[label=e3]{andrea.cerasa@ec.europa.eu}\orcid{0000-0002-2749-9142}}
\and
\author[C]{\fnms{Domenico}~\snm{Perrotta}\ead[label=e4]{domenico.perrotta@ec.europa.eu}\orcid{0000-0002-0987-3823}}
\address[A]{Department of Economics and Statistics, University of Siena, piazza San Francesco, 53100 Siena, Italy\printead[presep={,\ }]{e1}}

\address[B]{Department of Economics and Management, University of Parma, Via J.F. Kennedy 6, 43125 Parma, Italy\printead[presep={,\ }]{e2}}

\address[C]{European Commission, Joint Research Centre (JRC), Via E. Fermi 2479, 21027 Ispra, Italy\printead[presep={,\ }]{e3,e4}}
\end{aug}

\begin{abstract}
We address the task of identifying anomalous observations by analyzing digits under the lens of Benford’s law. Motivated by the statistical analysis of customs declarations, we answer one major and still open question: How can we detect the behavior of operators who are aware of the prevalence of the Benford's pattern in the digits of regular observations and try to manipulate their data in such a way that the same pattern also holds after data fabrication? This challenge arises from the ability of highly skilled and strategically minded manipulators in key organizational positions or criminal networks to exploit statistical knowledge and evade detection. For this purpose, we write a specific contamination model for digits, investigate distributional results on the fractional part of the significand and derive appropriate goodness-of-fit statistics for the considered adversarial testing problem. Along our path, we also unveil the peculiar relationship between two simple conformance tests based on the distribution of the first digit. We show the empirical properties of the proposed tests through a simulation exercise and application to real data from international trade transactions. Although we cannot claim that our results are able to anticipate data fabrication with certainty, they surely point to situations where more substantial controls are needed. Furthermore, our work can reinforce trust in data integrity in many critical domains where mathematically informed misconduct is suspected.
\end{abstract}


\end{frontmatter}



\section{Introduction}
\label{sec:intro}


Investigators in law-enforcement services, as well as auditors in insurance companies, financial institutions, and public bodies, are concerned with intentional manipulations of recorded financial transactions, driven not only by the prospect of economic gain, but also by the potential for individuals in key positions to falsify data for personal, political, or organizational motives. 
Benford’s law has become a widely used tool for detecting this type of anomalies in numerical data. However, its popularity has led to a new class of challenges: skilled individuals or organizations intent on deceiving investigators can manipulate fabricated data to mimic Benford’s distribution, thereby evading traditional detection methods. In this work, we address this vulnerability by developing statistical tests that remain effective under such adversarial scenarios. The proposed tests are exact, robust to mild contamination unrelated to fabrication, and capable of exposing deliberate attempts to mask manipulation. These advances significantly strengthen the reliability of Benford-based detection tools, reinforcing trust in data integrity across a range of domains including - but not limited to - the analysis of customs declarations, which serves as our primary application focus.

Our main motivation comes from the important application area 
of the statistical analysis of customs declarations, an established research track of the Joint Research Centre (JRC) of the European Commission stemming from the commitment to fight against fraud for the defense of the budget of the European Union (EU). This domain raises challenges that cannot be easily solved by statistical learning methods for fraud detection 
\citep{BRVV:2023}.
One reason 
is that the number of detected violations is typically very low when compared to the total number of transactions, leading to severe class imbalance problems. Second, while information on detected illegal activities can be trusted almost with certainty, great uncertainty should be attached to the remaining class, since investigations only cover a small minority of the available transactions and many infringements can go undetected. Last, but not least, the availability of data on certified illegal activities after the final court judgment is sometimes so much deferred in time that learning rules likely become outdated: therefore, advancements in anti-fraud approaches have to face with concurrent adaptations of the fraudulent practices.
Although some partial remedies for some of these issues have been recently outlined in other financial contexts \citep{dev+al:23,van+al:23}, the most effective methods 
conceive illegal customs declarations as anomalous transactions and identify them through formal outlier detection rules
\citep{cer:10,per+al:20,rou+al:19}.
%

In this work we address the task of identifying fabricated declarations in customs data through the perspective of analyzing transaction digits instead of transaction values. Anomaly detection in the space of transaction digits can be a powerful companion to more classical outlier detection tools, as proven by 
%
\cite{bar+al:16b,bar+al:21b,bar+al:21a}.
The approach that we adopt relies on the availability of a suitable model for the digits of genuine transactions, i.e. of transactions that originate from regular trade flows. 
%
\cite{cer+al:19} have shown that Benford's law can 
provide such a model under fairly general and easily verifiable conditions of trade, related to the ratio between the number of transactions and the number of traded goods of the subject under investigation. 
Furthermore, an important bonus of the digit approach is its ability to pinpoint serial misconduct, by focusing on the whole amount of transactions made by a single operator instead of looking at the potential anomaly of each transaction within the reference market, as done 
by the methods 
that deal with transaction values 
\citep{per+al:20}. 
The spirit of anomaly detection in the space of transaction digits is then similar to the recent approach followed by \cite{mag+al:24}, who focus on detecting the collective presence of outliers even if their precise identification may be challenging due to the weakness of individual signals.

%
As highlighted at the outset, the main goal of the present work is to answer one major and still open question: 
\begin{quote}
how can we detect the behavior of economic operators who are aware of the prevalence of the Benford's pattern in the digits of regular transactions and try to manipulate their data in such a way that the same pattern also holds after data fabrication?
\end{quote}
Such a challenging manipulation scheme undermines the available methodologies for fraud identification based on Benford's law.  
\citet[p. 11]{lac:19} emphasizes this potential threat in the specific context of the analysis of international trade data, while similar concerns exist in other financial scenarios \citep[p. 108]{kos:15}. More generally, the possibility that fraudsters adapt their approaches to state-of-the art analytical methods is known as ``concept drift'' in the machine learning literature 
and is now recognized as a crucial 
open issue in most disciplines 
\citep{BRVV:2023}.
One important task is thus to tackle the consequences of this subtle adversarial attack by unveiling malicious Benford-savvy behavior.
%
Although our focus is on the detection of infringements in international trade, our work should also be of interest in all the mathematically informed scenarios where lack of conformance to Benford's law is advocated as a signal of potential misconduct. These include accounting and finance \citep{nig:12,bar+sch:16,liu+al:23,ens+al:25}, electoral processes \citep{meb:10,per+tor:11,kli+al:12,gra+lac:18}, economics and national accounts \citep{tod:09,mic+sto:13,has+hos:19,wan+al:22}, on-line reviews \citep{ZW:2023}, and even the delicate field of evaluation of integrity of scientific research \citep{hor+al:20,eck+rux:23}.

We achieve our stated goal by means of four main 
developments.
First, we formalize the Benford-savvy 
malicious 
behavior that we aim to contrast through a new contamination model for digits, baptized the ``manipulated-Benford'' scheme, which extends the one presented in \cite{cer+al:19} to the Benford-savvy context. We then study the distributional properties of the fractional part of the significand, 
which is the most informative random quantity under the manipulated-Benford model. This study opens the door to the construction of new nonparametric tests of the Benford hypothesis which are able to unmask the postulated 
adversarial 
Benford-savvy behavior. Our third contribution is a general result which unveils the 
relationship between two available statistics for testing conformance to Benford's law, the first-digit chi-square statistic  \citep{nig:12} and the Hotelling statistic proposed in \cite{bar+al:21b}.
The nature of this relationship turns out to be very peculiar and amenable to a surprising simplification. It then leads us to suggest a 
new test statistic which is again effective to detect data fabrication under the manipulated-Benford model. Since none of 
our new tests can be expected to dominate the others under all possible configurations of the manipulated-Benford model, we finally combine them into an exact test of conformance to Benford's law which proves to be powerful under 
various specifications of this contamination scheme. 

We investigate the empirical properties of the proposed tests through 
a simulation exercise and 
application to real customs data. The former is designed to depict 
different 
possible specifications of the manipulated-Benford model and shows that 
our new tests can greatly outperform the existing ones, including the likelihood-ratio test 
of 
\cite{bar+al:23} which is instead optimal under local alternatives to Benford's law. We then represent the application potential of our approach in the case of 
suspicious economic operators analyzed by investigators
in a member state of the 
EU. The values declared by these traders do conform to Benford's law when only their first digit is considered, but they are picked as highly suspicious by our tests assuming a manipulated-Benford scheme. Although we cannot claim that our results are able to anticipate data manipulation with certainty, they surely point to situations where more substantial controls are needed in view of a possible serial and mathematically-informed illicit behavior. 

A final applied contribution of our work consists in a more realistic and robust Monte Carlo scheme for estimating the exact \textit{p}-values of the test statistics. The validity of Benford's law implies that transaction values are generated from a continuous random variable. This assumption is tenable under idealized trade conditions
\citep{cer+al:19}, but it might be mildly violated in practice due to the possible effects of truncation and rounding. The main consequence is that some of the transaction values may be recorded with a limited number of digits, even in the absence of substantial irregularities. Although our simulation exercise proves that only extreme approximations strongly affect inferential conclusions, we also propose a refined simulation scheme for computing the exact distribution function of each test statistic, given the observed behavior of the economic operator in terms of truncation (or rounding) of values. The advantage of our proposal, at the expense of a limited loss of power, is a much more accurate control of the false-positive rate under regular trade conditions that do not strictly adhere to Benford's law because of truncation or rounding problems.

The structure of the paper is as follows. In Section \ref{sec:Benford} we summarize the basic properties of Benford's law and describe the problem of testing conformance to it. Section \ref{sec:Frauds} defines the manipulated-Benford contamination model, through the concept of digit-wise contamination. 
in Sections \ref{sec:Frac} and \ref{sec:Properties} we derive the properties of the random quantities of interest under the manipulated-Benford model, obtain 
results on first-digit tests and develop a battery of 
tests of conformance specifically targeted to this Benford-savvy alternative. We then propose to combine our 
tests in Section \ref{sec:Comb} and validate them through a simulation exercise in Section \ref{sec:Pow}. Applications to real data from trade transactions are given in Section \ref{sec:Appl}.
Proofs are reported in the separate \textit{Supplementary Material}, together with 
some complementary results, both theoretical and empirical, our simulation algorithm for the truncated-Benford scheme and the description of a web application for anomaly detection in customs data through Benford's law.

\section{The Benford set-up}
\label{sec:Benford}

\subsection{Basic definitions and properties}
\label{sec:basic}

We define the ``significand function'' $S:\mathbb{R}\setminus\{0\}\rightarrow[1,10)$ as
\begin{equation*}
S(x)=10^{\langle\log_{10}|x|\rangle},
\end{equation*}
where $\langle x\rangle=x-\lfloor x\rfloor$ and $\lfloor x\rfloor=\max\{n\in\mathbb{Z}:n\leq x\}$ are the fractional part and the floor function of $x\in\mathbb{R}$, respectively. By assuming the probability space $(\Omega,\mathcal{F},P)$, an absolutely-continuous random variable $X$ is defined to be Benford 
\citep[p. 45]{ber+hil:15} 
if, for $u\in[1,10)$, the distribution function of $S(X)$ is 
\begin{equation}\label{UnifS}
    F_{S(X)}(u)=P(S(X)\leq u)=\log_{10}u .
\end{equation}
The first significant digit of $x$, denoted as $D(x)$, can be rephrased in terms of the significand function since $D(x)=\lfloor S(x)\rfloor$, and similar relationships also hold for the subsequent significant digits. 

Two fundamental properties of Benford random variables stem from \eqref{UnifS}. The first one is the celebrated first-digit law, stating that
\begin{equation}\label{d1}
    p_d=P(D(X)=d)=\log_{10}\left(\frac{d+1}{d}\right)
\end{equation}
for $d=1,\ldots,9$. Under Benford's law the distribution of the first significant digit is thus far from being uniform, 
but follows the logarithmic-type distribution ruled by \eqref{d1}. Furthermore, for $d=1,\ldots,9$, we let
\begin{equation}\label{Z1d}
    Z_{1,d}=\boldsymbol{1}_{[d,d+1)}(S(X)),
\end{equation}
where $\boldsymbol{1}_B$ is the indicator function of the set $B$, so that the occurrence probability for first-digit $d$ is given by $p_d=\expectation[Z_{1,d}]$.

The second feature of Benford random variables is called sum-invariance and is even more amazing. By defining
\begin{equation}\label{Z2d}
    Z_{2,d}=S(X)\boldsymbol{1}_{[d,d+1)}(S(X))
\end{equation}
for $d=1,\ldots,9$, the (first-digit) sum-invariance property of Benford's law states that
\begin{equation}\label{suminv}
    \expectation[Z_{2,d}]=C ,
\end{equation}
where $C=\log_{10}e$. 
The validity of Benford's law thus implies that the expected value of the significand of $X$ 
when the first digit is $d$, i.e. the expectation given in \eqref{suminv}, does not depend on $d$ and is equal to $C$. We refer to \cite{ber+hil:15} for a comprehensive and authoritative account of these and other mathematical characterizations of the law, while \cite{bar+al:21b} provide a detailed investigation of the 
relationship between \eqref{d1} and \eqref{suminv}. 

Benford's law has been challenging mathematicians and practitioners for decades 
\citep{ber+hil:11b}. It has also attracted the interest of statisticians mainly thanks to the limit theorem presented by 
\cite{hil:95}, which motivates the adoption of the law as the digit-generating model in many real-world situations. Theoretical results on the relationship between Benford random variables and classical univariate models for $X$  
are increasingly studied \citep{dum+leu:08,ber+hil:15,bar+al:23}, 
while an empirical investigation of the accuracy of the Benford property of \eqref{d1}
in the specific domain of international trade is detailed in \cite{cer+al:19}. In that context, 
alternative ways are suggested to analyze digits when Benford's law does not hold.

\subsection{Tests of the Benford hypothesis}
\label{sec:test}
We define as ``Benford hypothesis'' the statement that a random sample $X_1,\ldots,X_n$ of observations is made of $n$ realizations of a Benford random variable $X$. In view of \eqref{UnifS}, the Benford hypothesis can be 
translated into the null hypothesis
\begin{equation}\label{H0}
    H_0:S(X)\overset{\mathcal{L}}{=}10^U ,
\end{equation}
where $U$ is a Uniform random variable on $[0,1)$. The assessment of \eqref{H0} 
is the main inferential target 
of the statistical investigations 
based on transaction digits described in 
Section \ref{sec:Appl}. 

Since the validity of Benford's law implies the first-digit properties displayed in \eqref{d1} and \eqref{suminv}, many popular tests of the Benford hypothesis are based on first-digit statistics. The rationale is that disagreement with either \eqref{d1} or \eqref{suminv} contradicts the validity of \eqref{H0}, provided that null distributions are obtained under $H_0$. 
The simplest and perhaps most adopted strategy for testing \eqref{H0} in applications \citep[see, e.g.,][and the references therein]{nig:12,bar+al:16b} is to compute the first-digit Pearson statistic
\[
    Q_1=\sum_{d=1}^9\frac{(n\widehat{p}_d-np_d)^2}{np_d} ,
\]
where $p_d$ is given by \eqref{d1}, while 
\[
    \widehat{p}_d=\frac{1}{n}\,\sum_{i=1}^n\boldsymbol{1}_{\{d\}}(D(X_i))
\]
is its sample estimate. 
It is convenient for the developments that follow to express 
$Q_1$ as a quadratic form. If we recall 
\eqref{Z1d} and let $Z_1=(Z_{1,1}\ldots,Z_{1,8})^{\transpose}$, $\mu_1=\expectation[Z_1]$, $\varSigma_1=\Var[Z_1]$, under $H_0$ we have
\begin{equation*}
    \mu_1=(p_1,\ldots,p_8)^{\transpose} 
\end{equation*}
and
\begin{equation*}
    \varSigma_1=(p_d\delta_{d,d'}-p_dp_{d'}) ,
\end{equation*}
where $\delta_{d,d'}$ is the Kronecker delta function. 
Therefore, 
\begin{equation}
\label{Q1}
    Q_1=n(\bar{Z_1}-\mu_1)^{\transpose}\varSigma_1^{-1}(\bar{Z_1}-\mu_1) ,
\end{equation}
where $\bar{Z}_1=(\bar{Z}_{1,1},\ldots,\bar{Z}_{1,8})^{\transpose}$ and
\begin{equation*}
    \bar{Z}_{1,d}=\frac{1}{n}\,\sum_{i=1}^n \boldsymbol{1}_{[d,d+1)}(S(X_i)). 
\end{equation*}

Another powerful first-digit tool for testing $H_0$ is the 
Hotelling statistic proposed by 
\cite{bar+al:21b} on the basis of the sum-invariance property 
of
\eqref{suminv}. If we 
recall 
\eqref{Z2d} and let $Z_2=(Z_{2,1},\ldots,Z_{2,9})^{\transpose}$, $\mu_2=\expectation[Z_2]$, $\varSigma_2=\Var[Z_2]$, it follows that under $H_0$
\begin{equation*}
    \mu_2=(C,\ldots,C)^{\transpose} 
\end{equation*}
and
\begin{equation*}
    \varSigma_2=(C(d+1/2)\delta_{d,d'}-C^2) .
\end{equation*}
The 
Hotelling statistic is thus given by the quadratic form
\begin{equation}
\label{Q2}
    Q_2=n(\bar{Z}_2-\mu_2)^{\transpose}\varSigma_2^{-1}(\bar{Z}_2-\mu_2) ,
\end{equation}
where $\bar{Z}_2=(\bar{Z}_{2,1},\ldots,\bar{Z}_{2,9})^{\transpose}$ and
\begin{equation*}
    \bar{Z}_{2,d}=\frac{1}{n}\,\sum_{i=1}^n S(X_i)\boldsymbol{1}_{[d,d+1)}(S(X_i)).
\end{equation*}
Even if it relies on the first-digit sum-invariance property 
given in
\eqref{suminv}, $Q_2$ takes into account part of the information provided by all the recorded digits, through the sample averages in $\bar{Z}_2$ and through their covariance structure. This is a major advantage against Benford-savvy manipulations that we further exploit in Section \ref{sec:Properties}.

Strictly speaking, both $Q_1$ and $Q_2$ would test weaker hypotheses than the Benford one, since neither \eqref{d1} nor \eqref{suminv} imply \eqref{UnifS}. 
A test statistic specifically tailored to assess \eqref{H0} is instead the Kolmogorov-Smirnov (two-sided) statistic defined as
\begin{equation}\label{KS}
    KS_1 = \max \left\{A_1,B_1\right\},
\end{equation}
where 
\[
    A_1 = \max_{1\leq i\leq n}\left(\frac i n - \log_{10}S_{(i)}\right)\, ,\,
    B_1 = \max_{1\leq i\leq n}\left(\log_{10}S_{(i)}-\frac{i-1}{n}\right)
\]
and $S_{(1)} \le \cdots \le S_{(n)}$ are the order statistics of the significands $S(X_1),\ldots,S(X_n)$.
A scale invariant version of $KS_1$ is the Kuiper statistic
\begin{equation}\label{KU}
    KU_1 = A_1 + B_1.
\end{equation}
Furthermore, 
\cite{bar+al:23} derive two locally optimal test statistics (i.e., a likelihood-ratio and a score statistic) through a semiparametric approximation to the distribution function of $\log_{10}S(X)$. These statistics are scale invariant and improve over both $KS_1$ and $KU_1$ under standard departures from \eqref{H0}, when the law of $X$ is a member of large classes of distributions of wide applicability. 
We refer to \cite{bar+al:23} for a precise definition of such statistics and a description of their properties, as well as to \cite{Cerasa:22} for 
other discrepancy measures of potential interest in very small samples. 
We also note that in principle a myriad of alternative statistics could be inspired by the ideas of goodness-of-fit testing for directional  
data
\citep{cue+al:09,gar-por+al:20}, but they are not considered here.

\section{Manipulations that mimic Benford's law}
\label{sec:Frauds}

All the test statistics described in Section 
\ref{sec:test} are (to varying degrees) effective under ``regular'' alternatives to the Benford hypothesis, 
i.e.\ alternatives where all or a fraction of the random variables $X_1, \ldots, X_n$ are sampled from a distribution different from the Benford one. This setting can be formalized through a contamination model assuming that the distribution function $F_{S(X)}$ of the significand is an element within the following family of distribution functions 
\begin{equation}\label{contmodel}
    F_{S(X)} = (1-\lambda)F_{\mathrm{G}} + \lambda F_{\mathrm{C}},
\end{equation}
where $\lambda \in [0,1]$. In such a model, $F_{\mathrm{G}}$ is the distribution function of the ``good'' part of the data, i.e.\ $F_{\mathrm{G}}$ represents the postulated null model, $F_{\mathrm{C}}$ is the contaminant distribution, which is left unspecified except at most for the assumption of some regularity conditions, and $\lambda$ is the contamination rate. 
With emphasis on the first digit, \cite{cer+al:19} show that in the context of international trade it may be plausible to assume the validity of Benford's law under fairly general and easily verifiable conditions of trade, related to the ratio between the number of transactions $n$ and the number of traded goods, say $m$. Specifically, both $n$ and $m$ should be large enough, as anticipated by the fundamental limit theorem of \cite{hil:95}, and 
it is suggested to take $m \geq 0.2n$ as a practically sensible rule of thumb.
Under the conditions that ensure the validity of Benford's law, it thus follows from \eqref{UnifS} that we can take
\begin{equation}\label{benford}
    F_{\mathrm{G}}(u) = \log_{10}u 
\end{equation}
in
\eqref{contmodel}, with $u\in[1,10)$. 
\cite{bar+al:21b} and \cite{bar+al:23} 
provide extensive empirical evidence of the performance of many tests of the Benford hypothesis under \eqref{contmodel} and \eqref{benford}, when $\lambda=1$ and the contaminant distribution $F_{\mathrm{C}}$ corresponds to models for $X$ of popular choice 
in applications. 
Furthermore, the power of $Q_1$ in some cases where $\lambda<1$ is shown in \cite{cer+al:19}.

The Benford-savvy manipulation that we target encompasses digit-wise contamination and replaces the first digit of a fabricated value with the first digit of the realization of a Benford random variable. 
This is arguably the simplest and most attractive way to make up data following the Benford pattern for at least two major reasons. The first one is the existence of legal and administrative constraints (e.g., for balance sheets or customs declarations) that may prevent from producing a fully-fabricated sample of Benford values. The second motivation is the increasing popularity of the first-digit tests introduced in Section 
\ref{sec:test}, to which dishonest economic operators may try to adapt. 
If some of them were willing to fill administrative documents with fake data for the purpose of achieving an economic gain, the feeling that these data might be scrutinized for conformance with \eqref{d1} could lead to select first digits actually following Benford's law. Indeed, the development of new data manipulation schemes in response to existing fraud detection techniques is now recognized as a general and widespread challenge 
in most applications \citep{BRVV:2023,gal+al:24}.
%
It is apparent that the contamination model described by \eqref{contmodel} is not suitable to represent 
digit-wise departure from the Benford hypothesis and must then be 
replaced by a new one. 

We recall 
\eqref{UnifS} and define a random variable $X_{\mathrm{A}}$ such that
\[
    X_{\mathrm{A}}\overset{\mathcal{L}}{=} 10^U , 
\]
where $U$ is a Uniform random variable on $[0,1)$. Therefore, $X_{\mathrm{A}}$ is a Benford random variable and the probability distribution of $D(X_{\mathrm{A}})$ follows the first-digit law given in \eqref{d1}. We also consider a 
random variable $X_{\mathrm{B}}$ independent of $X_{\mathrm{A}}$. The Benford-savvy manipulated response is then
\begin{equation}\label{manip}
    S(X) = D(X_{\mathrm{A}}) + \langle S(X_{\mathrm{B}})\rangle .
\end{equation}
We define \eqref{manip} as the (first-digit) ``manipulated-Benford'' contamination model.

It is worth remarking that the manipulated-Benford model does not 
yield 
\eqref{UnifS} when $X_{\mathrm{B}}$ is itself a Benford random variable. More details on this, perhaps surprising, result are given in Section \ref{sec:Frac}, but intuition can anticipate that the manipulation devised in \eqref{manip} destroys the dependence structure among the digits of a Benford random variable. We also note that contamination models similar to the manipulated-Benford one are becoming of prominent interest for anomaly detection in other branches of statistical research. We refer to the framework of cell-wise contamination in multivariate estimation problems 
\citep{ray+rou:23} and we leave the study of its relationship with \eqref{manip} to further research.

%
%
%
%
%
%

\section{Tests for the manipulated-Benford alternative}
\label{sec:Frac}

Our goal consists in developing 
tests of the Benford hypothesis that can detect departures from it when the subtle manipulated-Benford alternative holds. It is clear from \eqref{manip} that statistics solely involving the first digit $D(X)$, such as $Q_1$ in 
\eqref{Q1}, cannot be effective for this purpose. 
Empirical investigation of the behavior of many existing statistics under model \eqref{manip} is deferred to Section \ref{sec:Pow}, where details about notation are provided. Nevertheless, we anticipate that even the most powerful scale-invariant test 
developed by 
\cite{bar+al:23} is barely effective against such a contamination, as shown in Table \ref{tab:lrt} for a few selected (and rather extreme) choices of distributions of $X_{\mathrm{B}}$ in \eqref{manip} and for moderately large sample sizes.

A 
direct approach to detect violations of the Benford hypothesis ruled by \eqref{manip} is to base inferences on the fractional part of $S(X)$, whose main distributional properties are given in the results that follow. 
Their proof is reported in the \textit{Supplementary Material}.

\begin{table*}[!t]
\caption{Estimated power, based on 5000 simulations, of the likelihood-ratio test of \cite{bar+al:23}, there denoted as $\varLambda_{\widehat{N},n}$, when $X$ follows the manipulated-Benford contamination model for some (rather extreme) choices of distributions of $X_{\mathrm{B}}$ in \eqref{manip} and for different sample sizes. The exact test size is 0.01. \label{tab:lrt}
}
\begin{center}
\begin{tabular}{ccccc}
\hline
$n$ & $\mathrm{Lognormal}(0.3,1)$ & $\mathrm{Weibull}(3.4,1)$ & $\mathrm{Uniform}[0, 1)$ & $\mathrm{Generalized \; Benford}(3)$  \\ 
\hline
200 & 0.013    & 0.014    & 0.022  & 0.036  \\
500 & 0.017    & 0.020    & 0.044  & 0.084  \\
1000& 0.023    & 0.023    & 0.082  & 0.205  \\
\end{tabular}
\end{center}
\end{table*}

\begin{proposition}
\label{propDF}
Assume that 
\eqref{H0} holds. 
For $v\in[1,10)$ and $u\in[0,1)$, 
the distribution function of the bivariate random vector $(D(X),\langle S(X) \rangle)^{\transpose}$ is
\[
    F_{(D(X),\langle S(X)\rangle)}(v,u)=\sum_{j=1}^{\lfloor v\rfloor}\log_{10}\left(\frac{j+u}{j}\right) .
\]
Moreover, 
for $u\in[0,1)$, 
the distribution function of $\langle S(X) \rangle$ is
\[
    F_{\langle S(X) \rangle}(u) = \sum_{d=1}^9\log_{10}\left(\frac{d+u}{d}\right) ,
\]
with corresponding probability density function
\[
    f_{\langle S(X) \rangle}(u) =\sum_{d=1}^9\frac{C}{d+u} .
\]
Finally, 
for $u\in[0,1)$, 
the conditional distribution function of 
$\langle S(X)\rangle$ given ${\{D(X)=d\}}$ 
is
\[
    F_{\langle S(X)\rangle\mid\{D(X)=d\}}(u)=\frac{1}{p_d}\log_{10}\left(\frac{d+u}{d}\right), 
\]
with corresponding probability density function
\[
f_{\langle S(X)\rangle\mid\{D(X)=d\}}(u)=\frac{C}{p_d(d+u)}.
\]
\end{proposition}

Proposition \ref{propDF} opens the door to the construction of 
nonparametric tests of the Benford hypothesis, which can be expected to be especially effective under the manipulated-Benford alternative. As in \eqref{KS} and \eqref{KU}, we focus both on the Kolmogorov-Smirnov statistic 
\begin{equation}\label{KSfr}
    KS_2 = \max\left\{A_2,B_2\right\}
\end{equation}
and on the Kuiper statistic
\begin{equation}
	\label{KUfr}
KU_2 = A_2 + B_2,
\end{equation}
where 
\[
    A_2 = \max_{1\leq i\leq n}\left(\frac i n - F_{\langle S(X) \rangle}(S_{2,(i)}) \right) \,,\,
    B_2 = \max_{1\leq i\leq n}\left(F_{\langle S(X) \rangle}(S_{2,(i)}) -\frac{i-1}{n}\right) ,
\]
and $S_{2,(1)} \le \cdots \le S_{2,(n)}$ represent the order statistics of the significand fractional parts $\langle S(X_1) \rangle,\ldots,\langle S(X_n) \rangle$. 
%

\begin{proposition}
\label{propE}
Under the same assumptions of Proposition \ref{propDF}, for $r,s\in\mathbb{N}$, we have
\begin{eqnarray*}
    \expectation[D(X)^r\langle S(X)\rangle^s] &=& (-1)^s\expectation[D(X)^{r+s}] + C \sum_{d=1}^9d^{r+s} \nonumber\sum_{j=1}^s\binom{s}{j}(-1)^{s-j}\frac{(1+1/d)^j-1}{j}.
\end{eqnarray*}
\end{proposition}

Proposition \ref{propE} yields the mixed moments of the joint distribution of $D(X)$ and $\langle S(X) \rangle$ when $X$ is a Benford random variable. In particular, after some algebra deferred to 
the \textit{Supplementary Material}, where its formal expression is given, 
the numerical value of the correlation coefficient between $D(X)$ and $\langle S(X)\rangle$ turns out to be
\begin{equation}
\label{cor}
    \cor[D(X),\langle S(X) \rangle] \simeq 0.05636 .
\end{equation}
We also obtain that the conditional expectation
\begin{equation*}
    \expectation[\langle S(X)\rangle\mid\{D(X)=d\}]=\frac{C}{p_d}-d
\end{equation*}
is nearly constant, since it takes values close to $0.5$ for all $d=1,\ldots,9$. Both measures thus show limited dependence between $D(X)$ and $\langle S(X) \rangle$ under 
\eqref{H0}.

As an interesting corollary of Proposition \ref{propE}, 
\eqref{cor} also explains why the manipulated-Benford contamination model defined in \eqref{manip} does not lead to the Benford hypothesis when $X_{\mathrm{B}}$ is itself a Benford random variable. Indeed, this choice of $X_{\mathrm{B}}$ yields $\cor[D(X_{\mathrm{A}}),\langle S(X_{\mathrm{B}})\rangle]=0$, $X_{\mathrm{A}}$ and $X_{\mathrm{B}}$ being independent, and contradicts \eqref{cor}.

\section{Properties of first-digit tests and another new test}
\label{sec:Properties} 

\subsection{The joint null distribution of $Q_1$ and $Q_2$}
\label{sec:joint}

We provide 
new insight on the properties of the first-digit test statistics $Q_1$ and $Q_2$ under the Benford hypothesis. 
Although a good Monte Carlo approximation to their exact joint distribution is available under \eqref{H0}, as described in Section \ref{sec:Pow},  
the result presented below gives their asymptotic joint null distribution. A major advantage of this large-sample result is that it suggests a very simple approximation 
which in Section 
\ref{sec:simpl} leads to a novel test statistic specifically tailored to the manipulated-Benford alternative. 

Let $Z=\mathrm{vec}(Z_1,Z_2)$, $\mu=\expectation[Z]$, $\varSigma_{12}=\text{Cov}[Z_1,Z_2]$ and $\varSigma=\Var[Z]$. 
If \eqref{H0} holds, 
\begin{equation*}
    \mu=\mathrm{vec}(\mu_1,\mu_2) 
\end{equation*}
and 
\begin{equation*}
    \varSigma=
    \begin{pmatrix}
    \varSigma_1&\varSigma_{12}\\
    \varSigma_{12}^{\transpose}&\varSigma_2 
    \end{pmatrix},
\end{equation*}
where $\mu_1$, $\mu_2$, $\varSigma_1$, $\varSigma_2$ are given in Section 
\ref{sec:test}. In addition, since
\begin{equation*}
    \expectation[Z_{1,d}Z_{2,d'}]=\delta_{d,d'}\expectation[Z_{2,d}]=C\delta_{d,d'},
\end{equation*}
it also follows that $\varSigma_{12}=(C\delta_{d,d'}-Cp_d)$. Correspondingly, if $\bar{Z}=\mathrm{vec}(\bar{Z}_1,\bar{Z}_2)$, we have $\expectation[\bar{Z}]=\mu$ and 
$\Var[\bar{Z}]=n^{-1}\varSigma$.
%
The following general result,
which is proved in the \textit{Supplementary Material},
leads to the large-sample distribution of the bivariate random vector $Q=(Q_1,Q_2)^{\transpose}$.

\begin{theorem}
\label{prop1}
Let $(Y_n)_{n\geq 1}$ be a sequence of centered random vectors defined on the probability space $(\Omega,\mathcal{F},P)$, such that $Y_n=\mathrm{vec}(Y_{1,n},Y_{2,n})$, where $Y_{1,n}$ and $Y_{2,n}$ are random vectors of dimensions $r_1$ and $r_2$, with $r_1\leq r_2$. Moreover, 
assume that 
\begin{equation*}
    \varPsi=\Var[Y_n]=
    \begin{pmatrix}
    \varPsi_1&\varPsi_{12}\\
    \varPsi_{12}^{\transpose}&\varPsi_2
    \end{pmatrix}
\end{equation*}
is a full-rank matrix with finite entries and that $Y_n$ converges in distribution to the random vector $Y=\mathrm{vec}(Y_1,Y_2)$ with the Normal law $N_{r_1+r_2}(0,\varPsi)$. If $(S_n)_{n\geq 1}$ is a sequence of bivariate random vectors such that $S_n=(S_{1,n},S_{2,n})^{\transpose}$ with $S_{1,n}=Y_{1,n}^{\transpose}\varPsi_1^{-1}Y_{1,n}$ and $S_{2,n}=Y_{2,n}^{\transpose}\varPsi_2^{-1}Y_{2,n}$, then $S_n$ converges in distribution to the bivariate random vector 
\[
    V=(V_1,V_2)^{\transpose}=(Y_1^{\transpose}\varPsi_1^{-1}Y_1,Y_2^{\transpose}\varPsi_2^{-1}Y_2)^{\transpose}
\]
with Laplace transform given by
%
\begin{eqnarray*}
    L_V(t_1,t_2)=(1+2t_2)^{-\frac{1}{2}(r_2-r_1)}\prod_{j=1}^{r_1}(1+2t_1+2t_2+4(1-\rho_j^2)t_1t_2)^{-\frac{1}{2}} ,
\end{eqnarray*}
where $1\geq\rho_1\geq\cdots\geq\rho_{r_1}\geq 0$ are the canonical correlations between $Y_1$ and $Y_2$, i.e. the singular values of the matrix $\varPsi_1^{-\frac{1}{2}}\varPsi_{12}\varPsi_2^{-\frac{1}{2}}$.
\end{theorem}

Theorem \ref{prop1} implies that $L_V(t_1,0)=(1+2t_1)^{-\frac{1}{2}r_1}$ and $L_V(0,t_2)=(1+2t_2)^{-\frac{1}{2}r_2}$, 
so that the marginal distributions of $V_1$ and $V_2$ are the $\chi_{r_1}^2$ and $\chi_{r_2}^2$ laws, respectively. 
Furthermore, since $\sqrt{n}(\bar{Z}-\expectation[\bar{Z}])$ converges in distribution to the Normal law $N_{17}(0,\varSigma)$,
by applying Proposition $1$ with $r_1=8$ and $r_2=9$ we obtain that $Q$ converges in distribution to the bivariate random vector $V$ with Laplace transform 
\begin{eqnarray*}
    L_V(t_1,t_2)=(1+2t_2)^{-\frac{1}{2}}\prod_{j=1}^8(1+2t_1+2t_2+4(1-\rho_j^2)t_1t_2)^{-\frac{1}{2}} .
\end{eqnarray*}
We thus have the stochastic representation $V\overset{\mathcal{L}}{=}\sum_{j=1}^9W_j$, 
where $W_j=(W_{1,j},W_{2,j})^{\transpose}$ are independent bivariate random vectors such 
that, for $j=1,\ldots,8$, each $W_j$ is distributed with 
Kibble's law of parameters $\rho_j$ and $\frac{1}{2}$ \citep{kib:41}, 
while $W_{1,9}$ is a Dirac random variable concentrated at zero and $W_{2,9}$ is distributed with the $\chi_1^2$ law. Kibble's law is actually a bivariate generalization of the Gamma distribution and is in turn a special case of the family discussed in \cite{tor+al:22}. Finally, by using algebraic software, we obtain
\begin{equation*}
    \expectation[V_1V_2]=\left.\frac{\partial L_V(t_1,t_2)}{\partial t_1\partial t_2}\right|_{t_1=0,t_2=0}=72+2\,\sum_{d=1}^8\rho_d^2 .
\end{equation*}
The correlation coefficient between $V_1$ and $V_2$ is then
\begin{equation*}
    \mathrm{cor}[V_1,V_2]=\frac{\sqrt{2}}{12}\,\sum_{j=1}^8\rho_j^2 ,
\end{equation*}
by recalling that the marginal distributions of $V_1$ and $V_2$ are the $\chi_8^2$ and $\chi_9^2$ laws, respectively. 

Since the entries of $\varSigma$ are known, the exact 
expression of 
$\rho_1,\ldots,\rho_8$ can 
be obtained by means of algebraic software and the corresponding numerical computation 
yields
the values reported in Table \ref{tab:cancor}, in such a way that $\cor[V_1,V_2]\simeq 0.9381$. Table \ref{tab:MCQ} displays the Monte Carlo estimates, based on $10^5$ replicates, of $\expectation[Q_1]$, $\expectation[Q_2]$, $\Var[Q_1]$, $\Var[Q_2]$ and $\cor[Q_1,Q_2]$ for selected values of $n$. It also compares these estimates with 
the moments of $V$, reached when $n \rightarrow \infty$. The agreement is remarkably close even in the case of small samples.

\begin{table*}[!t]
\caption{Numerical approximation to the canonical correlations between $Y_1$ and $Y_2$.\label{tab:cancor}
}
\begin{center}
\begin{tabular}{cccccccc}
\hline
$\rho_1$ & $\rho_2$ & $\rho_3$ & $\rho_4$ & $\rho_5$ & $\rho _6$ & $\rho_7$ & $\rho_8$ \\
\hline
0.9995   & 0.9994   & 0.9992   & 0.9990   & 0.9985   & 0.9977    & 0.9959   & 0.9906   \\
\hline
\end{tabular}
\end{center}
\end{table*}

\begin{table}[!t]
\caption{Monte Carlo estimates of some moments of $Q$, based on $10^5$ replicates, for selected sample sizes and comparison with asymptotic values.\label{tab:MCQ}
}
\begin{center}
\begin{tabular}{cccccc}
\hline
$n$&$30$&$50$&$100$&$200$&$\infty$\\
\hline
$\expectation[Q_1]$&$7.984$&$7.989$&$8.011$&$8.013$&$8$\\
$\expectation[Q_2]$&$8.979$&$8.999$&$9.019$&$9.014$&$9$\\
$\Var[Q_1]$&$16.538$&$16.366$&$16.466$&$16.413$&$16$\\
$\Var[Q_2]$&$18.355$&$18.336$&$18.190$&$18.167$&$18$\\
$\cor[Q_1,Q_2]$&$0.939$&$0.937$&$0.938$&$0.938$&$0.938$\\
\hline
\end{tabular}
\end{center}
\end{table}

\subsection{A simplified statistic}
\label{sec:simpl}

\begin{figure*}[!t]
\begin{center}
\includegraphics[width=0.45\textwidth]{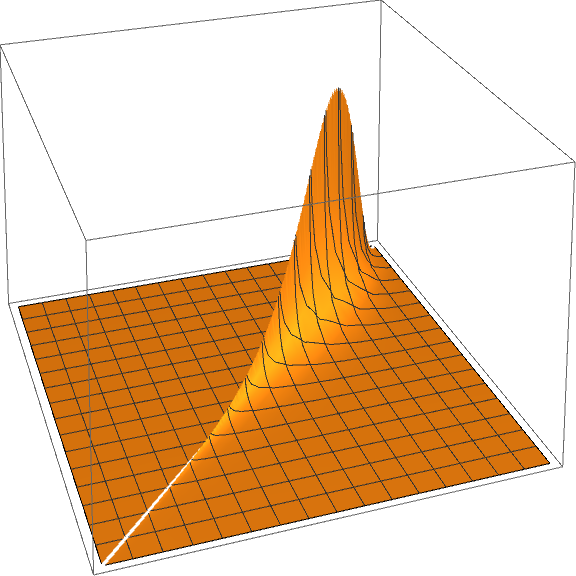}
\hfill
\includegraphics[width=0.45\textwidth]{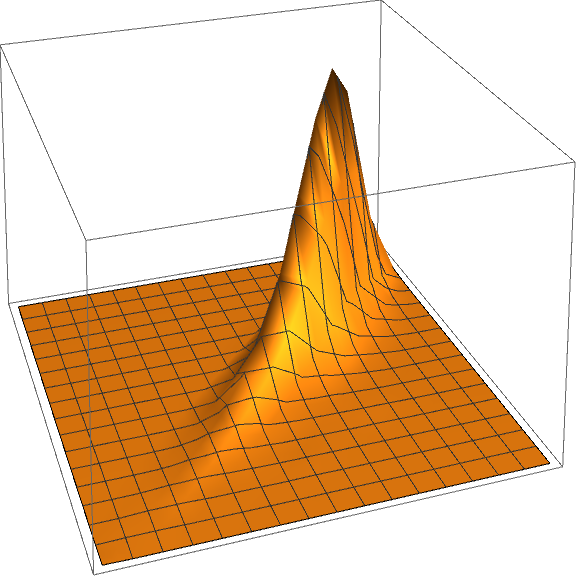}
\end{center}
\vspace{-0.25cm}
\caption{Probability density functions of $T$ (left-hand panel) and $V$ (right-hand panel).\label{fig:dens}}
\end{figure*}

The Laplace transform $L_V$ cannot be 
inverted analytically. 
However, the values of 
$\rho_1,\ldots,\rho_8$ 
displayed in Table \ref{tab:cancor}
suggest that a good approximation to the distribution of $V$ can be achieved by inverting $L_V$ under the assumption that $(\rho_1,\ldots,\rho_8)=(1,\ldots,1)$. Such an assumption gives rise to the bivariate probability density function of a new random vector, say
$T=(T_1,T_2)^{\transpose}$, such that
\begin{equation}\label{densT}
    f_T(x_1,x_2)=\frac{1}{96\sqrt{2\pi}}\,\frac{x_1^3}{\sqrt{x_2-x_1}}\,e^{-\frac{1}{2}x_2} ,
\end{equation}
for $x_1>0,x_2>x_1$.
%
The probability density function 
provided by 
\eqref{densT} is shown in the left-hand panel of Figure \ref{fig:dens}, while the right-hand panel displays the probability density function of $V$ obtained by numerical inversion. The agreement between the two plots is visually striking. 
Further peculiar features of \eqref{densT} are that the random variables $T_1$ and $(T_2-T_1)$ are independent under the Benford hypothesis and that $T_1$ is distributed with the $\chi_8^2$ law, while $(T_2-T_1)$ is distributed with the $\chi_1^2$ law. Therefore, this approximation keeps the same $\chi_8^2$ and $\chi_9^2$ laws for the marginal distributions, while the corresponding correlation becomes $\cor[T_1,T_2]=\frac{2\sqrt{2}}{3}\simeq 0.9428$, which is very close to the value of $\cor[V_1,V_2]$. 


The approximation to the asymptotic joint null distribution of $Q_1$ and $Q_2$ offered by \eqref{densT} 
suggests that the difference statistic 
\begin{equation}\label{delta}
    Q_{\varDelta}=Q_2-Q_1
\end{equation}
can be adopted to assess the Benford hypothesis under the manipulated-Benford alternative. The rationale is that the information provided by $(T_2-T_1)$, being independent of that of $T_1$, does not look at the first digit but at the subsequent ones. The same intuition also applies to the original test statistics $Q_1$ and $Q_2$, since computation of $Q_2$ involves all the significant digits through $\bar{Z}_2$ and the dispersion matrix $\varSigma_2$, while $Q_1$ is solely a function of the first digit.

\begin{figure*}[!t]
\begin{center}
\includegraphics[width=0.45\textwidth]{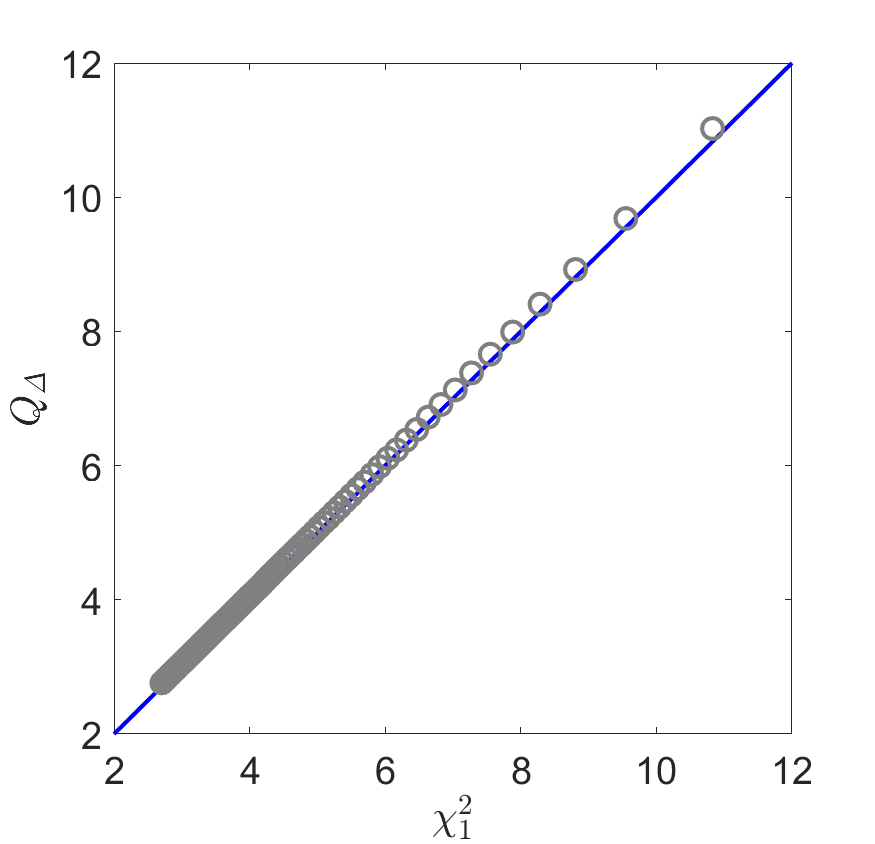}
\includegraphics[width=0.45\textwidth]{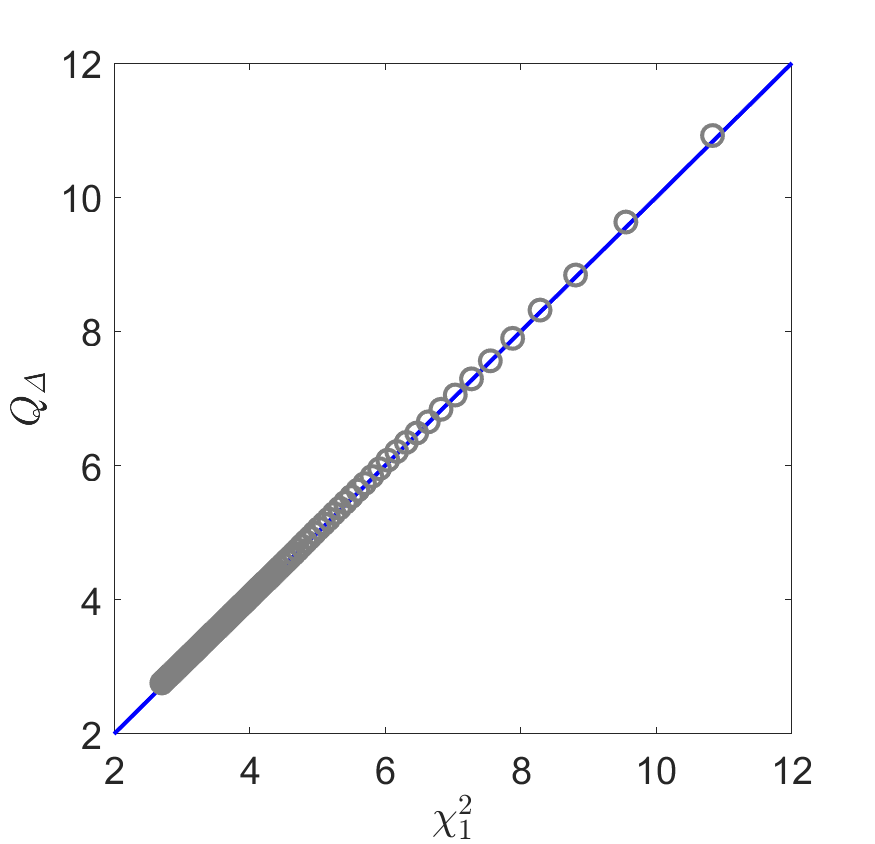}
\end{center}
\vspace{-0.5cm}
\caption{QQ-plot comparing the largest quantiles of $Q_{\varDelta}$ under \eqref{H0} with those of a $\chi^2_1$ random variable for $n=100$ (left-hand panel) and $n=500$ (right-hand panel). The plotted quantiles belong to the 10\% right-hand tail of each distribution.\label{fig:Delta}}
\end{figure*}

Asymptotically, we approximate the quantiles in the right-hand tail of the null distribution of $Q_{\varDelta}$ through those of the $\chi^2_1$ law, as it holds for its approximated counterpart $(T_2-T_1)$. Although we cannot expect this approximation to be valid across the whole support of $Q_{\varDelta}$, as it happens that $\{Q_1>Q_2\}$ and hence $\{Q_{\varDelta}<0\}$ with non-null probability, the fit of the $\chi^2_1$ distribution is amazingly good for testing purposes. For instance, Figure \ref{fig:Delta} contrasts the quantiles in the 10\% right-hand tail of the distribution of $Q_{\varDelta}$ under the Benford hypothesis, estimated through the Monte Carlo algorithm described in Section \ref{sec:Pow} and $10^6$ simulations, with those of a $\chi^2_1$ random variable. The match between the two sets of quantiles is very accurate even when $n=100$.


\begin{figure*}[!t]
\begin{center}
\includegraphics[width=0.45\textwidth]{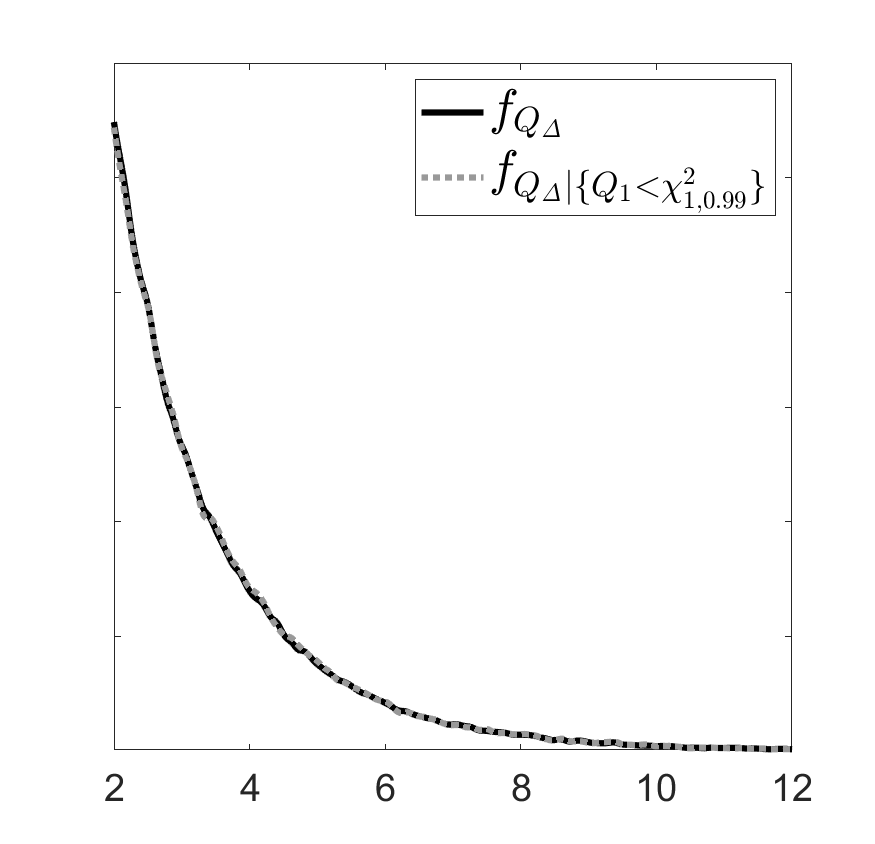}
\includegraphics[width=0.45\textwidth]{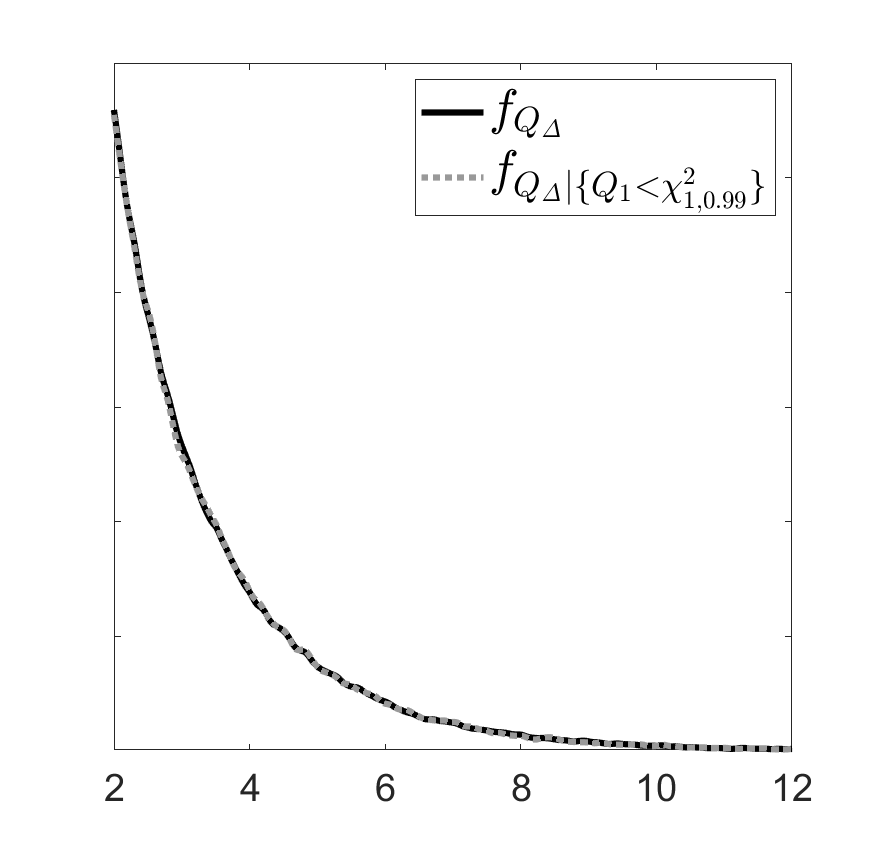}
\end{center}
\vspace{-0.5cm}
\caption{Zoom in the right-hand tail of the estimated null probability density of $Q_{\varDelta}$ (black solid line) and of the estimated null conditional probability density of $Q_{\varDelta}$, given that $Q_1 < \chi^2_{1,0.99}$ (gray dotted line). Left-hand panel: $n=100$; Right-hand panel: $n=500$. Estimates based on Matlab function \texttt{ksdensity} and $10^6$ simulated samples under $H_0$.\label{fig:DeltaBIS}}
\end{figure*}

Independence between $(T_2-T_1)$ and $T_1$ is also closely mirrored by $Q_{\varDelta}$ and $Q_1$, when we look at the right-hand tail of their distributions under the Benford hypothesis. Figure \ref{fig:DeltaBIS} compares a zoom of the estimated null probability density of $Q_{\varDelta}$, say $f_{Q_{\varDelta}}$, with the same portion of the estimated null conditional probability density of $Q_{\varDelta}$, given that $Q_1$ is below the 0.99 quantile of $\chi^2_1$, denoted as $f_{Q_{\varDelta}|\{Q_1<\chi^2_{1,0.99}\}}$, again for $n=100$ and $n=500$. The plotted kernel estimates are practically indistinguishable and show that the outcome of the first-digit test based on $Q_1$ has virtually no impact on the right-hand tail of the null distribution of $Q_{\varDelta}$. This result supports computation of $Q_{\varDelta}$ when there is no evidence of data fabrication in the first digit. In that case the observed value of the test statistic can be safely compared with the tail quantiles of its marginal distribution, in contrast to what happens when performing sequential tests on multiple digits and more complicated conditional distributions must be taken into account \citep{bar+al:16b}.


\section{Combination of tests}
\label{sec:Comb}

Since we want to leave the distribution of $X_{\mathrm{B}}$ in \eqref{manip} unspecified, none of our 
tests on the fractional part of $S(X)$ can be expected to dominate the others under all possible specifications of the manipulated-Benford model. Our proposal is then to consider a combined test statistic of \eqref{H0} tailored to detect the existence of Benford-savvy contamination under a 
variety of choices for $X_{\mathrm{B}}$ in \eqref{manip}. 

For concreteness, 
we focus on the combination of the Kolmogorov-Smirnov statistic 
given in \eqref{KSfr} with $Q_{\varDelta}$. 
The choice of a suitable combining function $\phi:[0,1]^2\rightarrow\mathbb{R}$ yields the test statistic
\begin{equation}\label{GKS}
    G_{KS}=\phi(1-F_{KS_2}(KS_2),1-F_{Q_{\varDelta}}(Q_{\varDelta})) ,
\end{equation}
where $F_{KS_2}$ and $F_{Q_{\varDelta}}$ are the distribution functions of $KS_2$ and $Q_{\varDelta}$, respectively, under the Benford hypothesis. 
Specifically, we adopt the minimum \textit{p}-value principle and let $\phi(u,v) = \min(u,v)$, so that the combined test rejects for small realizations of $G_{KS}$. The corresponding \textit{p}-value is then $p_{G_{KS}}(g)=F_{G_{KS}}(g)$, where $g$ is a realization of $G_{KS}$ and $F_{G_{KS}}$ is the distribution function of $G_{KS}$ under \eqref{H0}. 
Distribution function $F_{G_{KS}}$ can be approximated to an arbitrary degree of accuracy through the simple, but computationally efficient, Monte Carlo approach relying on \eqref{sim} and through its more robust variant to be described in the sections below. 
Therefore, a valid combined test is obtained through the estimate of $p_{G_{KS}}(g)$ and no further multiplicity correction is needed.
%
A similar test statistic is obtained by replacing $KS_2$ with the Kuiper statistic 
of \eqref{KUfr}. We 
denote as $G_{KU}$ the analogue of \eqref{GKS} combining $KU_2$ with $Q_{\varDelta}$.

\section{Simulation study}
\label{sec:Pow}

\subsection{Aims of the study}

In this section we provide simulation evidence of the empirical advantages of our 
tests under the manipulated-Benford contamination model. Specifically, we compare the performance of \eqref{KSfr}, \eqref{KUfr} and \eqref{delta}, as well as of their combinations, to that of the existing tests introduced in Section \ref{sec:Benford}\ref{sec:test}. We include in our comparison also the two-digit version of the Pearson statistic, 
here written as $Q_{12}$, and the likelihood-ratio statistic 
of \cite{bar+al:23}, denoted as $\varLambda_{\widehat{N},n}$. The former is often advocated to test \eqref{H0} in moderately large samples, as it extends consideration beyond the first digit 
\citep[pp. 78--80]{nig:12}, while the latter is a scale-invariant statistic which enjoys the optimality properties of a likelihood approach.

We emphasize that we cannot expect our 
statistics $KS_2$, $KU_2$ and $Q_{\varDelta}$ to be particularly effective under the standard contamination scheme 
defined in 
\eqref{contmodel}, since they neglect information on the first digit $D(X)$. 
We argue that such a limitation does hinder the usefulness of our proposal, the main ability of $KS_2$, $KU_2$ and $Q_{\varDelta}$ being to detect data fabrication in the fractional part of $S(X)$ when $D(X)$ indeed follows \eqref{d1}. 
Nevertheless, our tests can be also combined with any first-digit test to detect departures from the Benford hypothesis in the absence of a specific suspicion of Benford-savvy manipulations. The advantage of this approach, 
whose investigation is left for further research, is to obtain simple procedures with good power against both the naive manipulations which agree with \eqref{contmodel} and the more subtle data fabrication schemes following \eqref{manip}, without the need to 
select the type of departure from $H_0$ in advance.

In our power comparisons, we rely on Monte Carlo estimates of the exact quantiles of each test statistic. For this purpose, $B$ Monte Carlo replicates of the test statistic under consideration, say $T=T(X_1,\ldots,X_n)$, are generated under the Benford hypothesis as
\begin{equation}\label{sim}
	T^\dagger_b = T(10^{U_{b,1}},\ldots,10^{U_{b,n}}) , 
\end{equation}
for $b = 1,\ldots,B$, where $U_{b,1},\ldots,U_{b,n}$ are independent Uniform random variables on $[0, 1)$. For a realization $t$ of $T$, the Monte Carlo estimates of the exact \textit{p}-values are then computed as $p^\dagger_T(t) = 1 - F^\dagger_T(t)$, where 
\[
    F^\dagger_T(t) = \frac{1}{B}\sum_{b=1}^B \boldsymbol{1}_{(-\infty,t]}(T^\dagger_b).
\]
The estimates of some relevant quantiles of our 
test statistics are given in 
the \textit{Supplementary Material} using $B=10^6$ replicates. 

In the simulations that follow we take $\gamma=0.01$ to be the exact test size. Results obtained with different exact quantiles from \eqref{sim} are qualitatively very similar. 
We only show power results for $n=500$, while those for different sample sizes are given in 
the \textit{Supplementary Material}, together with further details about the test statistics not described in Section
\ref{sec:test} and the adopted simulation algorithm (including the link to the code that can be used to replicate our simulations). 
Power comparisons are performed using 5000 independent runs of the manipulated-Benford model 
of
\eqref{manip}, for different selections of $X_{\mathrm{B}}$. 

\subsection{Evidence under the manipulated-Benford model}

Our choices for $X_{\mathrm{B}}$ in the manipulated-Benford model are displayed in Table \ref{tab:n500}. The first two scenarios, the Lognormal and the Weibull laws, follow \cite{bar+al:21b} and assume that the contaminant random variable $X_{\mathrm{B}}$ comes from distributional models which are popular in many fields, including the analysis of economic aggregates that arise in trade \citep{bar+al:16a}. In both instances, we  
denote by $\alpha$ the shape parameter of the law and take the scale parameter equal to 1. We focus on the effect of $\alpha$ since it is known that Lognormal and Weibull random variables become practically indistinguishable from a Benford one for suitable choices of their shape parameter \citep{dum+leu:08,ber+hil:15}. Our subsequent scenario assumes that $X_{\mathrm{B}}$ is a Uniform random variable defined on interval $[0,\alpha)$. This setting can be taken as a representative of ``human'' data fabrication, as it is often argued that number invention by humans might be biased towards simple distributions such as the Uniform law \citep[see][pp. 300--306 for some examples of human number invention]{nig:12}. Here we are not interested in the effect of $\alpha$ itself, but we consider different values of the range to represent our lack of knowledge about the support chosen for the purpose of data invention. 
Our final power scenario widens the perspective by considering the possibility that data manipulation directly involves a model for the digits of $X$. We thus assume that $X$ is a Generalized Benford random variable with parameter $\alpha\in \mathbb{R}$. 
The Generalized Benford model has been introduced to represent digit distributions when $X$ follows a power-law \citep{pie+al:01,bar+pra:20}. For instance, the first-digit masses span from $p_1 = 0.556$ to $p_9 = 0.012$ when $\alpha=-1$, while we recover the Uniform first-digit distribution $p_1 = \ldots = p_9 = 1/9$ when $\alpha=1$. 
The resulting expressions for $F_{S(X)}$ and $F_{\langle S(X) \rangle}$ are reported in 
the \textit{Supplementary Material}.

The manipulated-Benford scenario is particularly challenging for all the test statistics derived under 
\eqref{contmodel}. Taking into account the full significand $S(X)$, as is done by $KS_1$, $KU_1$ and $\varLambda_{\widehat{N},n}$, does not help in any instance. Only the mixed approach provided by $Q_2$ (and to a much lesser extent by $Q_{12}$) is able to keep power to a reasonable level, a result anticipated 
in \cite{bar+al:21b} for the case of a Generalized Benford random variable. Nevertheless, our new tests clearly outperform $Q_2$, as well as the others competitors, in all the selected specifications of \eqref{manip}. It is also seen that the ordering among $KS_2$, $KU_2$ and $Q_{\varDelta}$ can change according to the specific contaminant distribution, but that the power of $G_{KS}$ and $G_{KU}$ is always close to that of the best performer. We thus conclude that our new tests are to be generally recommended for the purpose of detecting Benford-savvy manipulations and that the loss of power induced by combination of tests is minor, especially if compared to the resulting increased robustness against different specifications of the nature of the manipulating digit distribution.

\begin{table*}[!t]
	\caption{Estimated power, based on 5000 simulations, under the manipulated-Benford model for different choices of $X_{\mathrm{B}}$ in \eqref{manip}, when $n=500$. The exact test size is $\gamma=0.01$. Null quantiles of test statistics are estimated with $B=10^6$ replicates in \eqref{sim}.
}
		\label{tab:n500}
\begin{small}
\begin{center}
		\begin{tabular}{@{}cccccccccccc@{}}
	\hline
	$\alpha$ & $Q_1$ & $Q_{12}$ & $Q_2$ & $KS_1$ & $\varLambda_{\widehat{N},n}$ & $KU_1$ & $KS_2$ & $KU_2$ & $Q_{\varDelta}$ & $G_{KS}$ & $G_{KU}$   \\ 
	\hline
	\multicolumn{12}{c}{$\mathrm{Lognormal}(\alpha,1)$}\\
    0.3   & 0.011 & 0.451 & 0.630 & 0.034 & 0.017 & 0.038 & 0.997 & 0.986 & 0.927 & 0.996 & 0.981 \\
    0.4   & 0.010 & 0.096 & 0.122 & 0.015 & 0.010 & 0.014 & 0.759 & 0.575 & 0.381 & 0.702 & 0.528 \\
    0.5   & 0.009 & 0.033 & 0.021 & 0.014 & 0.010 & 0.013 & 0.304 & 0.167 & 0.076 & 0.241 & 0.135 \\
    0.6   & 0.010 & 0.023 & 0.011 & 0.013 & 0.011 & 0.015 & 0.105 & 0.051 & 0.018 & 0.079 & 0.039 \\
	\hline		
	\multicolumn{12}{c}{$\mathrm{Weibull}(\alpha,1)$}\\
    2.2   & 0.010 & 0.040 & 0.034 & 0.016 & 0.011 & 0.015 & 0.426 & 0.259 & 0.123 & 0.365 & 0.216 \\
    2.6   & 0.010 & 0.104 & 0.130 & 0.016 & 0.011 & 0.014 & 0.819 & 0.661 & 0.406 & 0.771 & 0.615 \\
    3.0   & 0.010 & 0.246 & 0.319 & 0.027 & 0.016 & 0.025 & 0.970 & 0.922 & 0.721 & 0.959 & 0.895 \\
    3.4   & 0.010 & 0.466 & 0.533 & 0.036 & 0.020 & 0.043 & 0.997 & 0.991 & 0.883 & 0.996 & 0.986 \\
	\hline
	\multicolumn{12}{c}{$\mathrm{Uniform}[0, \alpha)$}\\
    5     & 0.010 & 0.047 & 0.234 & 0.015 & 0.041 & 0.050 & 0.364 & 0.211 & 0.614 & 0.564 & 0.535 \\
    20    & 0.009 & 0.042 & 0.235 & 0.019 & 0.039 & 0.047 & 0.360 & 0.203 & 0.598 & 0.559 & 0.530 \\
    40    & 0.010 & 0.044 & 0.235 & 0.018 & 0.045 & 0.047 & 0.368 & 0.199 & 0.600 & 0.554 & 0.533 \\
    60    & 0.010 & 0.047 & 0.238 & 0.018 & 0.040 & 0.047 & 0.353 & 0.200 & 0.604 & 0.557 & 0.533 \\		
	\hline
	\multicolumn{12}{c}{$\mathrm{Generalized \; Benford}(\alpha)$}\\
    -1.0  & 0.011 & 0.153 & 0.225 & 0.018 & 0.011 & 0.016 & 0.884 & 0.722 & 0.589 & 0.852 & 0.703 \\
    1.0   & 0.010 & 0.048 & 0.250 & 0.018 & 0.045 & 0.050 & 0.356 & 0.200 & 0.593 & 0.546 & 0.514 \\
    2.0   & 0.010 & 0.124 & 0.577 & 0.025 & 0.065 & 0.078 & 0.768 & 0.568 & 0.903 & 0.887 & 0.867 \\
    3.0   & 0.012 & 0.217 & 0.774 & 0.049 & 0.084 & 0.109 & 0.930 & 0.796 & 0.974 & 0.969 & 0.960 \\	
	\hline
\end{tabular}
\end{center}
\end{small}
\end{table*}

\subsection{Truncation and rounding}

We carry on with our simulation exercise by investigating the effect of contamination in the last significant digits, as it happens with rounding errors or other numerical inaccuracies. Although the Benford hypothesis assumes $S(X)$ to be a continuous random variable, rounding and truncation to a rather limited number of significant digits often occur in practice even in the absence of 
malicious data manipulation. We require our tests to be fairly robust against such violations of $H_0$ and need to devise an alternative practical strategy when it is not the case,
since rounding and inaccuracies typically have very different implications than data fabrication. 

\begin{figure*}[!t]
\begin{center}
\includegraphics[width=0.45\textwidth]{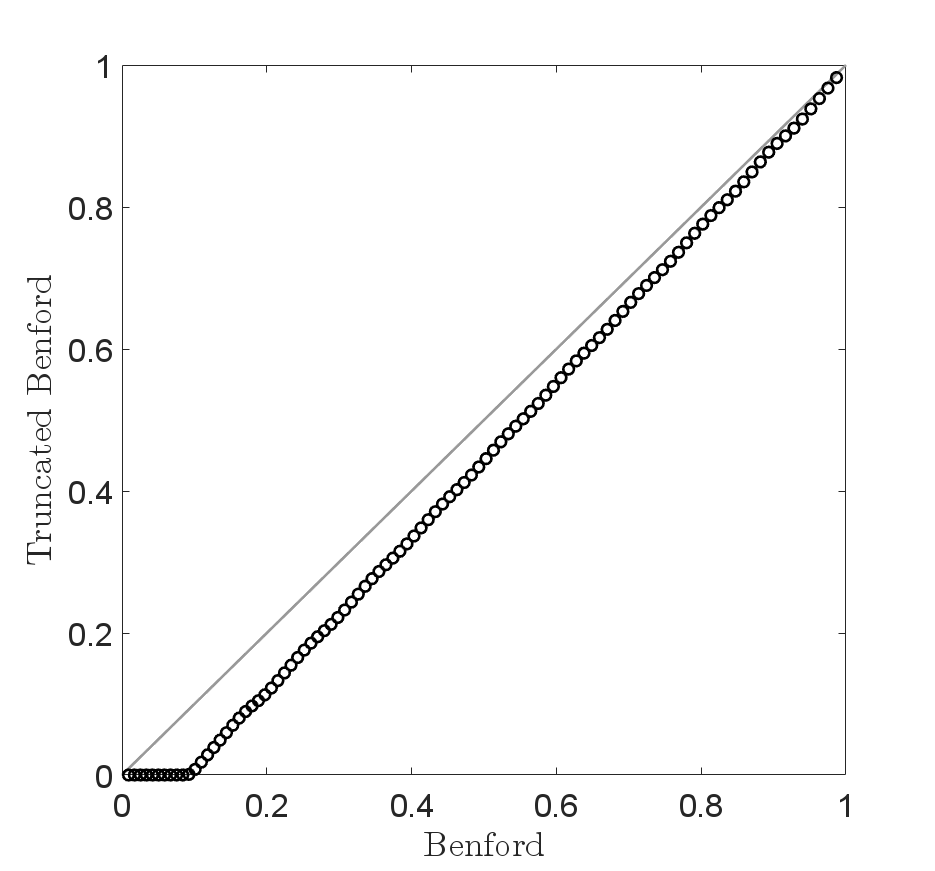}
\includegraphics[width=0.45\textwidth]{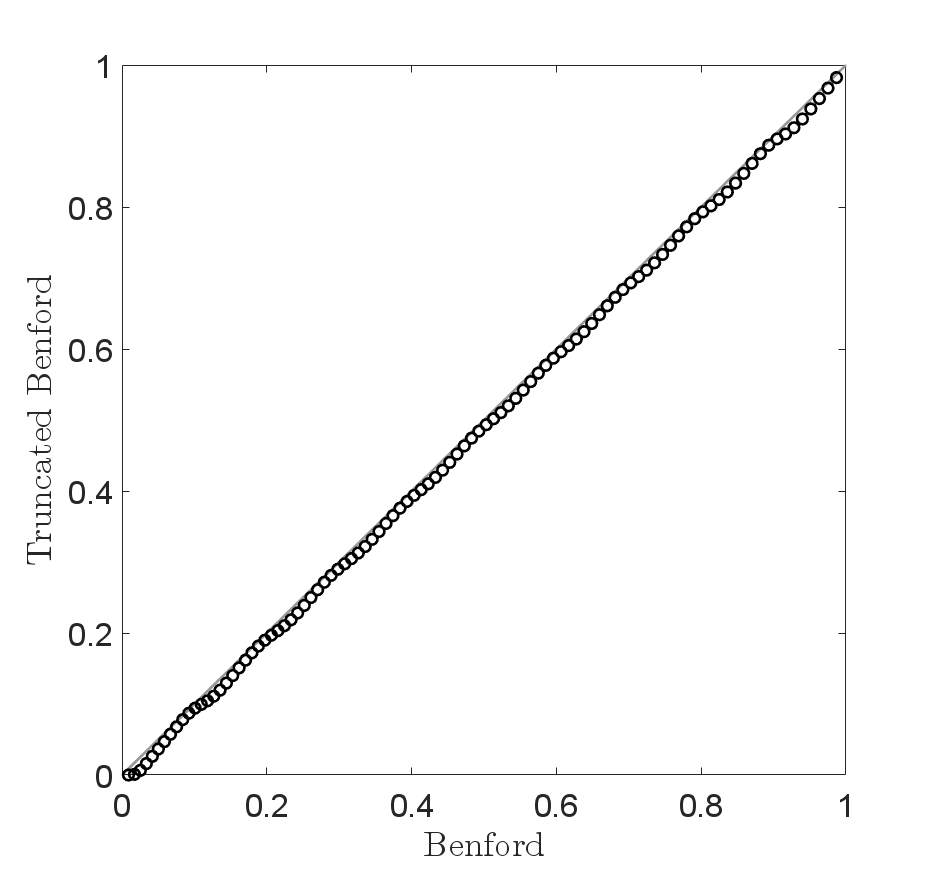}
\end{center}
\vspace{-0.5cm}
\caption{QQ-plots comparing the empirical distribution of $\langle S(X)\rangle$ under the Benford hypothesis with that obtained using the truncated-Benford algorithm described in the Supplementary Material, in two cases with $n=500$. Left: $n_1=n_2=50$ and $n_k=100$ for $k=3,\ldots,6$. Right: $n_1=0$ and $n_k=100$ for $k=2,\ldots,6$.\label{fig:n500Trunc}}
\end{figure*}

In Table \ref{tab:n500trunc} we compute the proportion of rejections of $H_0$ when the Benford significand $S(X)$, simulated according to \eqref{sim}, is truncated to have $k$ digits. We see that the effect of truncation is negligible, unless $k$ is very small. This result is perhaps surprising, since our tests look at the fractional part of $S(X)$ and the chosen simulation scheme depicts a worst-case scenario for them, by assuming that the same truncation level affects all the simulated significands.
Nevertheless, in Section \ref{sec:Appl} we apply a more robust simulation algorithm which takes truncation (or rounding) into account and allows us to control the exact size of the tests of \eqref{H0} for any specific truncation (or rounding) scheme. This robust simulation algorithm is detailed in 
the \textit{Supplementary Material}.
To convey the effect of discretization, Figure \ref{fig:n500Trunc} contrasts the quantiles of $\langle S(X)\rangle$ under the Benford hypothesis with those obtained using our truncated-Benford algorithm, in two instances where $n=500$ and $B=10^5$ replicates are used to estimate quantiles. If, for $k=1,\ldots,K$, we define $n_k$ to be the number of sample values with $k$ significant digits in their significand, the left-hand panel displays the case $n_1=n_2=50$ and $n_k=100$ for $k=3,\ldots,6$, while the right-hand panel is obtained with $n_1=0$ and $n_k=100$ for $k=2,\ldots,6$. It is seen that only extreme forms of truncation 
may be expected to have considerable impact on inferential conclusions based on $\langle S(X) \rangle$, while even a moderate fraction of sample values with $k=2$ (as in the right-hand panel) does not distort the qualitative impression of the distribution of $\langle S(X) \rangle$ gathered under the Benford hypothesis. However, we stress that replacing the Benford simulation scheme based on \eqref{sim} with the truncation-robust algorithm described in the Supplementary Material will provide estimates of exact quantiles and exact \textit{p}-values for any observed realization of $n_1,\ldots,n_K$, even in situations (like the one depicted in the left-hand panel of Figure \ref{fig:n500Trunc}) where wild truncation occurs without being related to deliberate data manipulation.
Additional empirical evidence is displayed in 
the \textit{Supplementary Material}, where we also cover the case of rounding.

\begin{table}[!t]
	\caption{Proportion of rejections of $H_0$, based on 5000 simulations, under a truncated-Benford model with $k$ significant digits in $S(X)$, when $n=500$. The exact test size is $\gamma=0.01$. Null quantiles are the same as in Table \ref{tab:n500}.
}
		\label{tab:n500trunc}
\begin{small}
\begin{center}
		\begin{tabular}{@{}cccccc@{}}
	\hline
	$k$ & $KS_2$ & $KU_2$ & $Q_{\varDelta}$ & $G_{KS}$ & $G_{KU}$   \\ 
	\hline
    6     & 0.009 & 0.010 & 0.009 & 0.008 & 0.009 \\
    5     & 0.010 & 0.011 & 0.010 & 0.010 & 0.010 \\
    4     & 0.011 & 0.009 & 0.010 & 0.011 & 0.011 \\
    3     & 0.020 & 0.024 & 0.016 & 0.016 & 0.021 \\
    2     & 1.000 & 1.000 & 0.756 & 1.000 & 1.000 \\	
	\hline
\end{tabular}
\end{center}
\end{small}
\end{table}



\section{Applications in international trade}
\label{sec:Appl}

That Benford's law has an almost 
immeasurable potential in the detection of various types of digit manipulations is becoming part of the collective imagination, with a rapidly increasing scientific literature 
%
\citep{ber+al:09,bar+al:21a} complemented by recent 
narrative and documentary production inspired from it 
\citep{net:20,mur:23}.
%
At the present time, it is then almost obvious to consider the possibility that dishonest operators in the international trade circuit -- like other economic,  financial or politically motivated networks disposing of analytical capabilities -- may consider having their movements scrutinized under Benford's law. 
It is recognized that the ability of fraudsters to modify their patterns in response to progresses in analytical research is one of the reasons why fraud has not yet disappeared from virtually all application domains \citep{BRVV:2023}.

%
The Joint Research Centre (JRC) of the European Commission has established a research track on international trade data and Benford's law both under its own institutional activities and alongside several Customs services through bilateral pilot activities.
The cases in this section are used 
primarily to illustrate the concrete applicability of our approach in this pilot phase.  
The declaration of the traded value at customs has direct financial consequences for the economic operator, who has to pay the duties, and for the public bodies collecting them, which in the 
EU
involve both the Member States and the European Institutions.  A dishonest operator has motivation to manipulate the value to minimize the taxable amount or to get more favorable treatments.  In the former case the typical amounts involved are relatively big and the price is under-declared.  The latter case can occur also with small amounts when involving tax exemption thresholds: for example, in the EU,  
the consignments that do not exceed 150 euros are not subject to duty payments 
and data with an over-representation of figures below such a threshold would be 
suspicious.  For an in-depth understanding of this multi-faced problem we refer to the 
monograph by \cite{giangiacomo:2023}, covering 
the main issues 
that affect the financial interests of the 
EU.

The data in question are the declarations made by the economic operators at import or export, which are very rich.  Customs services 
scrutinize all these variables when verifying the compliance of a declaration.  In our simplified experimentation,  based on data provided by a partner Customs service,  we just look at the declared value in euro and,  to distinguish different samples,  we use an anonymized identifier both for the operator and for the commodity code.      

%
%
%
%
%

\begin{figure*}[!t]
\begin{center}
\includegraphics[width=0.5\textwidth]{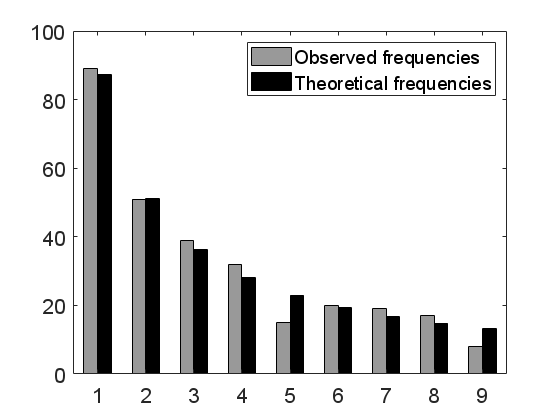}
\includegraphics[width=0.4\textwidth]{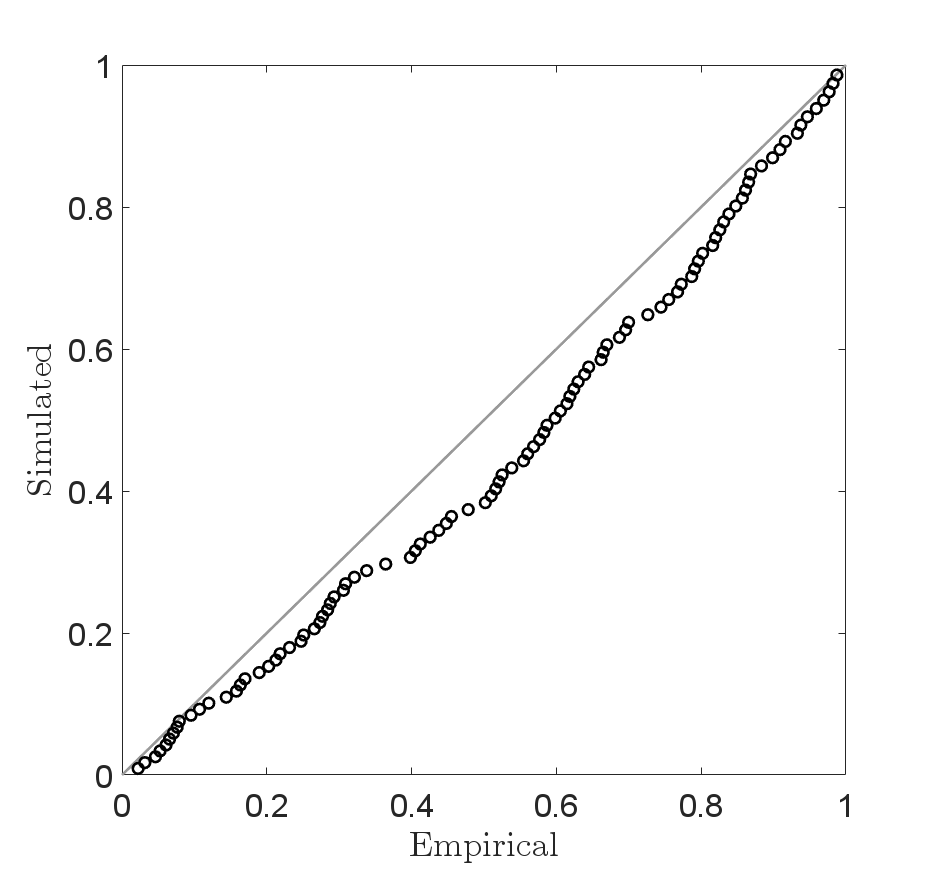}
\end{center}
\vspace{-0.5cm}
\caption{Left: Histogram comparing the first-digit distribution of Operator A with the theoretical frequencies computed under the Benford probabilities \eqref{d1}. Right: QQ-plot comparing the empirical quantiles of $\langle S(X)\rangle$ for the same operator with their Monte Carlo estimates under the truncated-Benford model.\label{fig:TraderA}}
\end{figure*}

In the customs declarations of the first economic operator that we analyze, say Operator A, we observe $n=290$ distinct values of $S(X)$ and only a limited degree of truncation, since $n_1=0$, $n_2=2$ and $n_3=5$. The conditions described 
in \cite{cer+al:19} for the validity of \eqref{benford} are broadly satisfied for this trader and indeed its empirical first-digit distribution, displayed in the left-hand panel of Figure \ref{fig:TraderA}, visually confirms the impression of close conformance to the theoretical counts predicted by \eqref{d1}. The corresponding exact \textit{p}-value of $Q_1$ is around 0.6, based on $B=10^5$ replicates of \eqref{sim}. Similarly, the exact \textit{p}-values of $Q_{12}$, $KS_1$ and $KU_1$ are close to 0.19, 0.25 and 0.10, respectively. None of these tests casts doubts on the validity of \eqref{H0}, a conclusion driven by the major effect of the empirical distribution of $D(X)$ on the test statistics. A preliminary signal of potential disagreement with the Benford hypothesis in the digits following the first one is provided by $Q_2$, whose exact \textit{p}-value boils down to less than 1\%. We then sharpen the conclusion suggested by the Hotelling statistic through a detailed analysis of the fractional part of the distinct significands available for this agent. The right-hand panel of Figure \ref{fig:TraderA} compares the estimated quantiles of $\langle S(X) \rangle$ to those obtained under the truncated-Benford algorithm introduced in Section \ref{sec:Pow} and detailed in the Supplementary Material, again based on $B=10^5$ simulations. Strong disagreement is now paramount. This qualitative finding is supplemented by the exact \textit{p}-values of our 
tests, which are all much smaller than 0.01 under the truncated-Benford model, and in particular by $Q_{\varDelta}$, whose \textit{p}-value is below 0.0001. Therefore, in contrast to the evidence provided by the standard statistics, our approach reinforces the idea that the values of $\langle S(X) \rangle$ do not follow the Benford hypothesis for this operator, even when the observed (mild) truncation is taken into account.


\begin{figure*}[!t]
\begin{center}
\includegraphics[width=0.5\textwidth]{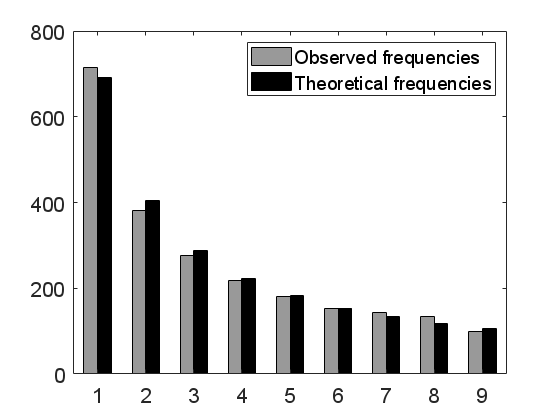}
\includegraphics[width=0.4\textwidth]{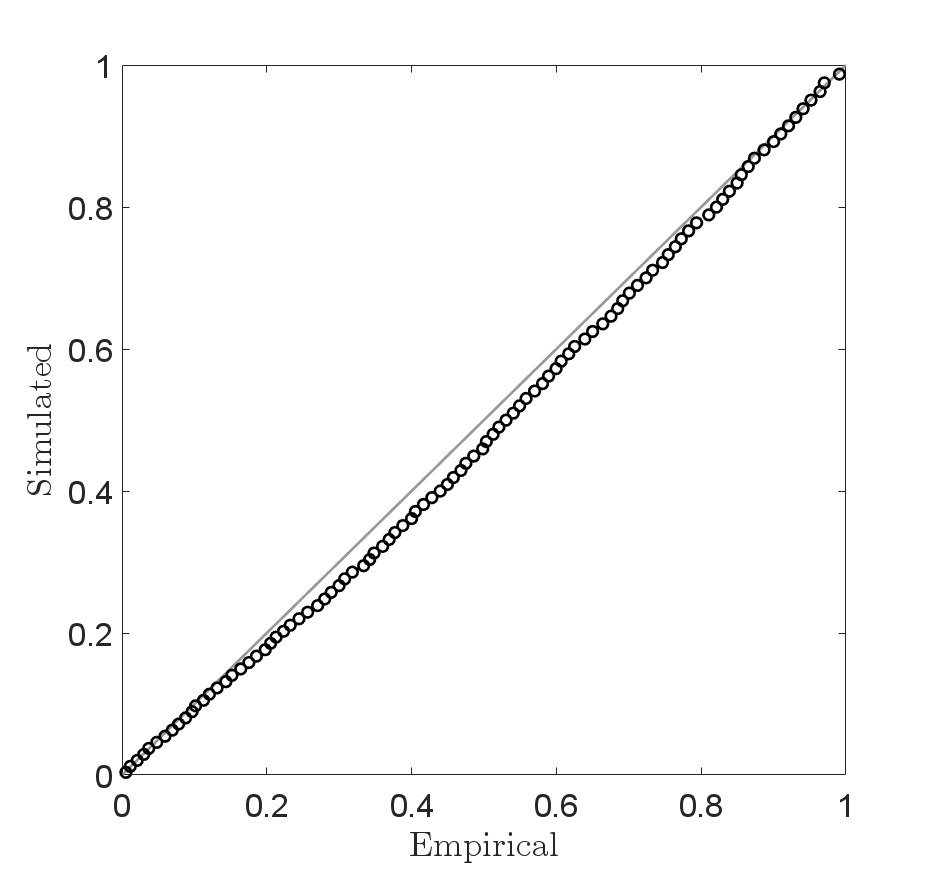}
\end{center}
\vspace{-0.5cm}
\caption{As Figure \ref{fig:TraderA}, but now for Operator B.\label{fig:TraderB}}
\end{figure*}

The situation is even more intriguing in the case of Operator B, 
an instance for which Benford's law might be considered a questionable model for the digits of trade flows in view of the findings of \cite{cer+al:19}, being $m \simeq 0.05n$. It is then 
puzzling to see that the observed distribution of $D(X)$, pictured in the left-hand panel of Figure \ref{fig:TraderB}, closely adheres to the theoretical Benford distribution. Visual inspection is confirmed by statistical analysis, since the asymptotic \textit{p}-value of the first digit statistic $Q_1$ is 0.58, based on a quite large sample of $n=2298$ distinct values of $S(X)$. 
Similar information is provided by the other statistics described in Section \ref{sec:Benford}\ref{sec:test}, whose \textit{p}-values are all higher than 0.15 after allowing for truncation through our algorithm in
the \textit{Supplementary Material} 
(with $n_1=9$, $n_2=39$, $n_3=100$ and $B=10^5$). None of the standard statistics, including $Q_2$, thus provides evidence against the Benford hypothesis for this agent, in spite of the fact that the number of traded goods is relatively low with respect to the number of transactions and the theoretical conditions established by \cite{hil:95} for the validity of \eqref{UnifS} are brought into question, especially in the case of a large sample. However, our conclusion becomes very different when looking at the fractional part of $S(X)$, displayed in the right-hand panel of Figure \ref{fig:TraderB}. The exact \textit{p}-values of $KS_2$ and $KU_2$ under the truncated-Benford scheme 
are both close to 0.002. The difference statistic $Q_{\varDelta}$ is less extreme (\textit{p}-value around 0.04), but $G_{KS}$ and $G_{KU}$ inherit much of the evidence provided by $\langle S(X)\rangle$, their exact \textit{p}-values being 0.004 and 0.003, respectively. The versatility and good power of these combined tests are not surprising in view of the results of the simulation study of Section \ref{sec:Pow}.


We have seen that our approach proves to be very informative in the two analyzed examples, 
but we cannot claim to be able to anticipate Benford-savvy manipulations with certainty, since unforeseen -- but licit -- economic and trading factors might possibly affect the significands of the financial flows under investigation. Nevertheless, our results surely point to situations where more substantial controls are needed and where comparison with the findings of other statistical techniques designed to signal possible financial infringements, such as those described 
in \cite{per+al:20}, could be particularly useful. 
Verification of the actual presence of a fraudulent Benford-tactic is in the hands of the authorities.

\begin{acks}[Acknowledgments]
The authors are grateful to Ivano Azzini for his skillful and patient assistance in interfacing the Fortran and MATLAB codes described in the Supplementary Material, an essential contribution to ensuring the computational reproducibility of this work. They also warmly thank Emmanuele Sordini for his dedication to the development of WebAriadne, a tool which makes our new tests both accessible and user-friendly for authorized analysts. The meticulous ongoing efforts  of these two collaborators are instrumental in promoting the broader dissemination and practical application of the methods in this study. 

The research line on Benford’s law presented here stems from the broader experience gained through longstanding efforts in support of EU policies related to international trade and specific anti-fraud initiatives. The ideas developed in this work have been inspired by, or built upon, insights from these wider research tracks. In this context, substantive contributions and ideas have also come from researchers at the Joint Research Centre and the University of Parma, with Marco Riani and Francesca Torti deserving special mention for their significant role in shaping and advancing this collective research agenda.

\end{acks}
\begin{funding}
The Joint Research Centre (JRC) of the European Commission is funded by Horizon Europe and the Euratom Research and Training Programme. JRC has studied the data sets and problems of the paper in the course of bilateral pilot activities with some Customs services of the European Union. 
The research of Andrea Cerioli has been supported by the Italian Project of Relevant National Interest (PRIN) n. 2022LANNKC, titled ``Innovative statistical tools for the analysis of large and heterogeneous customs data''. This PRIN program is financially supported by  NextGenerationEU (\url{https://next-generation-eu.europa.eu}).
\end{funding}

\begin{supplement}
\stitle{Proof, additional simulations, data and tools }
\sdescription{The supplements of the paper in appendix consist of several integrations, providing: 
the proofs of our theoretical results;
some ancillary theoretical derivations;
simulation results which complement those given in the text, together with further algorithmic details;
a description of the content of the GitHub repository, containing the code that can be used to replicate our simulations and the anonymized significands used in our applications;
our refined simulation scheme for computing the exact distribution function of each test statistic, given the observed behavior in terms of truncated or rounded values; the description of an enhanced web application, called WebAriadne, for anomaly detection in customs data through Benford's law.
}
\end{supplement}

\begin{supplement}
\stitle{GitHub repository: \url{https://github.com/AndreaCerioliUNIPR/Benford-savvy}}
\sdescription{
 The Benford-savvy GitHub repository  contains the code to replicate our simulations and the anonymized significands used in our applications. In particular, the anonymized and ordered significands of Operator A and B are available in the files \texttt{sig\_A.txt} and  \texttt{sig\_B.txt}.}
\end{supplement}


\bibliographystyle{imsart-nameyear} 
\bibliography{BCCP_main}


\newpage

\renewcommand{\appendixname}{Supplementary Material}

\appendix

\section{Technical Extensions}

%
This Supplementary Material  provides a technical extension to the main article by including:
\begin{itemize}
\item the proofs of our theoretical results;
\item ancillary theoretical derivations;
\item algorithmic details and a link to the code repository for replicating our simulations;
\item additional simulation results that complement those in the main text;
\item a refined simulation scheme for computing the exact distribution function of each test statistic, accounting for the impact of truncation or rounding in the observed data.
\end{itemize}

\subsection{Proof of Proposition 1}

If hypothesis (6) holds, $X$ is a Benford random variable. 
We thus have
\begin{equation*}
    \begin{aligned}[t]
    F_{(D(X),\langle S(X)\rangle)}(v,u)
    &=P(D(X)\leq v,\langle S(X)\rangle\leq u)=\sum_{j=1}^{\lfloor v\rfloor} P(j\leq S(X)\leq j+u)\\
    &=\sum_{j=1}^{\lfloor v\rfloor}(F_{S(X)}(j+u)-F_{S(X)}(j))=\sum_{j=1}^{\lfloor v\rfloor}\log_{10}\left(\frac{j+u}{j}\right),
    \end{aligned}
\end{equation*}
for $v\in[1,10)$ and $u\in[0,1)$.
Therefore, for $u\in[0,1)$, the distribution function of $\langle S(X)\rangle$ is
\begin{equation*}
    F_{\langle S(X)\rangle}(u)=P(D(X)\leq 9,\langle S(X)\rangle\leq u)=\sum_{d=1}^9\log_{10}\left(\frac{d+u}{d}\right),
\end{equation*}
while the expression of the probability density function $f_{\langle S(X) \rangle}$ follows on differentiation. Finally, the conditional distribution function of $\langle S(X)\rangle$ given $\{D(X)=d\}$ is 
\begin{equation*}
    \begin{aligned}[t]
    F_{\langle S(X)\rangle\mid\{D(X)=d\}}(u)
    &=\frac{P(S(X)=d,\langle S(X)\rangle\leq u)}{P(D(X)=d)}=\frac{P(d\leq S(X)\leq d+u)}{p_d}\\
    &=\frac{F_{S(X)}(d+u)-F_{S(X)}(d)}{p_d}=\frac{1}{p_d}\log_{10}\left(\frac{d+u}{d}\right), 
\end{aligned}
\end{equation*}
for $u\in[0,1)$. The expression of the probability density function $f_{\langle S(X)\rangle\mid\{D(X)=d\}}$ follows on differentiation.


\subsection{Proof of Proposition 2}

If $f_{S(X)}$ denotes the probability density function of $S(X)$, from (1) we have
\begin{equation*}
    f_{S(X)}(u)=\frac{C}{u} , 
\end{equation*}
for $u\in[1,10)$.
The case $s=0$ is obvious. For $s=1,2,\ldots$, we have
\begin{equation*}
    \begin{aligned}[t]
    \expectation[D(X)^r\langle S(X)\rangle^s]
    &=\int_1^{10}D(u)^r\langle S(u)\rangle^sf_{S(X)}(u)du=\sum_{d=1}^9\int_d^{d+1}D(u)^r\langle S(u)\rangle^sf_{S(X)}(u)du\\
    &=\sum_{d=1}^9\int_d^{d+1}d^r(u-d)^sf_{S(X)}(u)du.
    \end{aligned}
\end{equation*}
Thus, by means of the binomial theorem, 
we obtain
\begin{equation*}
    \expectation[D(X)^r\langle S(X)\rangle^s]=\sum_{d=1}^9d^r \int_d^{d+1} \phi_{d,s}(u)f_{S(X)}(u)du,
\end{equation*}
where
\begin{equation*}
    \phi_{d,s}(u)=(-d)^s+\sum_{j=1}^s\binom{s}{j}u^j(-d)^{s-j}
\end{equation*}
Since 
\begin{equation*}
\int_d^{d+1}f_{S(X)}(u)du=p_d\,,\,\int_d^{d+1}u^jf_{S(X)}(u)du=C\,\frac{(d+1)^j-d^j}{j}, 
\end{equation*}
we have
\begin{eqnarray*}
    \expectation[D(X)^r\langle S(X)\rangle^s]=(-1)^s\sum_{d=1}^9d^{r+s}p_d+C\sum_{d=1}^9d^{r+s}\sum_{j=1}^s\binom{s}{j}(-1)^{s-j}\frac{(d+1)^j-d^j}{jd^j}
\end{eqnarray*}
and the result follows.

\subsection{Proof of Theorem 1}
The components of the random vector $S_n$ are quadratic forms depending on $Y_{1,n}$ and $Y_{2,n}$, respectively. 
Under the assumptions, $S_n$ then converges in distribution to $V$ as $n\rightarrow\infty$ \citep[Corollary 1.7]{ser:80}. 
By 
the definition of canonical correlations, there exist two orthogonal matrices $\varGamma_1$ of order $(r_1\times r_1)$ and $\varGamma_2$ of order $(r_2\times r_2)$, such that
\begin{equation*}
    \varGamma_1\varPsi_1^{-\frac{1}{2}}\varPsi_{12}\varPsi_2^{-\frac{1}{2}}\varGamma_2=H ,
\end{equation*}
where $H=(D,0_{r_1\times(r_2-r_1)})$ is a matrix of order $(r_1\times r_2)$ with $D=\mathrm{diag}(\rho_1,\ldots,\rho_{r_1})$. Hence, by 
denoting $U=\mathrm{vec}(U_1,U_2)$ with
\begin{equation*}
    U_1=(U_{1,1},\ldots,U_{1,r_1})^{\transpose}=\varGamma_1\varPsi_1^{-\frac{1}{2}}Y_1
\end{equation*}
and
\begin{equation*}
    U_2=(U_{2,1},\ldots,U_{2,r_2})^{\transpose}=\varGamma_2\varPsi_2^{-\frac{1}{2}}Y_2 ,
\end{equation*}
we have
\begin{equation*}
    \Var[U]=
    \begin{pmatrix}
    I_{r_1}&H\\
    H^{\transpose}&I_{r_2}
    \end{pmatrix}.
\end{equation*}
The random vectors $U_1$ and $U_2$ are then distributed with the Normal laws $N_{r_1}(0,I_{r_1})$ and $N_{r_2}(0,I_{r_2})$, respectively, while 
$\Cov[U_1,U_2]=H$. In addition, we have 
\begin{equation*}
V_1=U_1^{\transpose}U_1=\sum_{j=1}^{r_1}W_{1,j}
\end{equation*}
and
\begin{equation*}
V_2=U_2^{\transpose}U_2=\sum_{j=1}^{r_2}W_{2,j} ,
\end{equation*}
where $W_{1,j}=U_{1,j}^2$ and $W_{2,j}=U_{2,j}^2$. Considering the expression of $\Var[U]$, for $j=1,\ldots,r_1$, the bivariate random vectors $W_j=(W_{1,j},W_{2,j})^{\transpose}$ are independent and in turn independent of the random variables $W_{2,r_1+1},\ldots,W_{2,r_2}$. 
Each $W_j$ is distributed according to the Kibble's law \citep{kib:41} of parameters $\rho_j$ and $\frac{1}{2}$, with Laplace transform 
\begin{eqnarray*}
    L_{W_j}(t_1,t_2)=\expectation[e^{-t_1W_{1,j}-t_2W_{2,j}}]=(1+2t_1+2t_2+4(1-\rho_j^2)t_1t_2)^{-\frac{1}{2}} ,
\end{eqnarray*}
while $W_{2,r_1+1},\ldots,W_{2,r_2}$ are independently distributed with the $\chi_1^2$ law. Therefore, the Laplace Transform of $V$ 
is
\begin{equation*}
    \begin{aligned}[t]
    L_V(t_1,t_2)&=\expectation[e^{-t_1V_1-t_2V_2}]=\expectation[e^{-t_1 \sum_{j=1}^{r_1}W_{1,j}-t_2\sum_{j=1}^{r_2}W_{2,j}}]\\
&=\prod_{j=r_1+1}^{r_2}\expectation[e^{-t_2W_{2,j}}]\,\prod_{j=1}^{r_1}\expectation[e^{-t_1W_{1,j}-t_2W_{2,j}}] ,
\end{aligned}
\end{equation*}
and the result follows.

\subsection{Ancillary theoretical derivations}
\subsubsection{Correlation between $D(X)$ and $\langle S(X) \rangle$}

We provide details on the derivation of the correlation coefficient between $D(X)$ and $\langle S(X)\rangle$ when $X$ is a Benford random variable. In the notation of the paper and using algebraic software, we have
\begin{equation*}
    \expectation[D(X)]=\sum_{d=1}^9 d\log_{10}\left(\frac{d+1}{d}\right)\simeq 3.44024
\end{equation*}
and
\begin{equation*}
    \expectation[D(X)^2]=\sum_{d=1}^9 d^2\log_{10}\left(\frac{d+1}{d}\right)\simeq 17.8917,
\end{equation*}
from which $\Var[D(X)]\simeq 6.05651$. In addition, 
\begin{equation*}
    \expectation[\langle S(X) \rangle]=-\expectation[D(X)]+9C\simeq 0.46841
\end{equation*}
and
\begin{equation*}
    \expectation[\langle S(X) \rangle^2]=\expectation[D(X)^2]-\frac{81}{2}\,C\simeq 0.30281 ,
\end{equation*}
from which $\Var[\langle S(X) \rangle]\simeq 0.08340$. Since 
\begin{equation*}
    \expectation[D(X)\langle S(X) \rangle]=- \expectation[D(X)^2]+45C \simeq 1.65151,
\end{equation*}
the correlation coefficient between $D(X)$ and $\langle S(X) \rangle$ under the Benford hypothesis is 
\begin{equation*}
    \begin{aligned}[t]
    \cor[D(X),\langle S(X) \rangle]&=\frac{45C-\Var [D(X)]-9C\expectation[D(X)]}{\sqrt{\Var [D(X)](\Var [D(X)]+18C\expectation[D(X)]-81C(C+1/2))}}\\
    &\simeq 0.05636 .
    \end{aligned}
\end{equation*}

\subsubsection{Distribution functions under the Generalized Benford model}

Let $X$ be a Generalized Benford random variable with parameter $\alpha\in \mathbb{R}$. 
In this case we have
\begin{equation*}
	F_{S(X)}(u)=
	\begin{cases}
	\vspace{0.15cm}  
	\log_{10}u \qquad &\alpha=0 \\
		\frac{u^\alpha-1}{10^\alpha-1} \qquad &\alpha\neq 0
		\vspace{0.25cm}
	\end{cases}
\end{equation*}
for $u\in[1,10)$.  
Furthermore,
\begin{equation*}
	F_{\langle S(X) \rangle}(u)=
	\begin{cases}
	\vspace{0.15cm}  
	\sum_{d=1}^9 \log_{10}(\frac{d+u}{d}) \qquad &\alpha=0 \\
		\frac{1}{10^\alpha-1}\sum_{d=1}^9((d+u)^\alpha - d^\alpha) \qquad &\alpha\neq 0
		\vspace{0.25cm} 
	\end{cases}
\end{equation*}
for $u\in[0,1)$.

%
%
%

\subsection{Algorithmic details and link to code}


The simulations in 
Section 6
of the paper have been performed through a Fortran code which extends the one 
available through the Supplementary Material of \cite{bar+al:23}. 
The simulations have been run on a computer with 64 GB of RAM, i9 processor and 64-bit Windows 
operating system.
Our Fortran program makes use of some IMSL functions and routines, including \texttt{DKSONE} for the computation of the Kolmogorov-Smirnov and Kuiper statistics. Since this routine does not admit ties, some mild form of jittering may be necessary under the models involving truncation or rounding. 
Our Fortran code is available at

\begin{center}
\url{https://github.com/AndreaCerioliUNIPR/Benford-savvy}
\end{center}

\noindent together with an almost complete mirror implementation in Matlab. A compatible version of the original Fortran code (i.e., a version without IMSL routines) can also be run from Matlab using a wrapper to the source file compiled using the mex function. 
This Matlab wrapper (created for 64-bit Windows operating system) is available in the same repository referenced above.

In addition to the test statistics defined in 
the paper, in our simulation study we provide comparison with the (exact) tests based on the following statistics:
\begin{itemize}
\item $Q_{12}$: two-digit Pearson statistic testing conformance to the first-two digit distribution implied by the Benford hypothesis \citep[see, e.g.,][]{nig:12,bar+al:16b}. This statistic is written as
\begin{equation}
\label{Q12}
Q_{12} = \sum_{d_1=1}^9 \sum_{d_2=0}^9\frac{(n\widehat{p}_{d_1,d_2}-np_{d_1,d_2})^2}{np_{d_1,d_2}} ,
\end{equation}
where $d_1 \in \{1,\ldots,9\}$ and $d_2 \in \{0,\ldots,9\}$, 
\[
p_{d_1,d_2} = \log_{10}\left(1+\frac{1}{10d_1+d_2}\right)
\]
is the probability that the first and the second digit of $X$ are equal to $d_1$ and $d_2$, respectively, when $X$ is a Benford random variable, and $\widehat{p}_{d_1,d_2}$ is the sample estimate of $p_{d_1,d_2}$.
\item $\varLambda_{\widehat{N},n}$: Likelihood Ratio statistic of the Benford hypothesis with data-driven selection of the number of non-negative trigonometric components through the BIC criterion \citep[see][Sections 4.3 and 4.4]{bar+al:23}.
\end{itemize}

\subsection{Complement to simulation results}

\subsubsection{Null simulations}

In Table \ref{tab:quant} of this Supplement we provide 
estimated quantiles of our new test statistics under the Benford hypothesis. These estimates are computed through the Monte Carlo algorithm described in 
Section 6
of the paper, using $B=10^6$ replicates and relying on the stochastic representation 
\begin{equation}
\label{benf}
X\overset{\mathcal{L}}{=}10^U,
\end{equation}
with $U$ a Uniform random variable on $[0,1)$. We also make use of the well known property
\begin{equation}
\label{significand}
S(10^U)=10^U .
\end{equation}


\begin{table*}[t]
\caption{Estimated quantiles of the proposed test statistics under the Benford hypothesis, based on $10^6$ replicates, for different sample sizes $n$ and for different tail probabilities $\gamma$.}
\label{tab:quant}
\begin{center}
	\begin{tabular}{@{}cccccc@{}}
	$n$ & $KS_2$ & $KU_2$ & $Q_{\varDelta}$ & $G_{KS}$ & $G_{KU}$   \\ 
	\hline
	\multicolumn{6}{c}{$\gamma=0.10$}\\
 200 & 0.086      & 0.113     & 2.768        & 0.064          & 0.056 \\
 500 & 0.054      & 0.072     & 2.771        & 0.064          & 0.056 \\
1000 & 0.038      & 0.051     & 2.766        & 0.064          & 0.056 \\
	\hline		
	\multicolumn{6}{c}{$\gamma=0.05$}\\
 200 & 0.095      & 0.122     & 3.893        & 0.031          & 0.027 \\
 500 & 0.060      & 0.078     & 3.900        & 0.031          & 0.027 \\
1000 & 0.043      & 0.055	    & 3.883        & 0.031          & 0.027 \\
	\hline
	\multicolumn{6}{c}{$\gamma=0.01$}\\
 200 & 0.114      & 0.140     & 6.683        & 0.006          & 0.005 \\
 500 & 0.072      & 0.089     & 6.689        & 0.006          & 0.005 \\
1000 & 0.051      & 0.063     & 6.681        & 0.006          & 0.005 \\
	\hline
\end{tabular}
\end{center}

\end{table*}


\subsubsection{Simulation results for different sample sizes}

Table \ref{tab:n200} and Table \ref{tab:n1000} of this Supplement provide power results under the manipulated-Benford alternative for $n=200$ and $n=1000$, respectively, under the same manipulation models considered in Table 3 of the paper. Table \ref{tab:trunc} of this Supplement instead investigates the effect of truncating $S(X)$ to have $k$ significant digits, again when $n=200$ and $n=1000$.

\begin{table*}[t]
	\caption{Estimated power, based on 5000 simulations, under the manipulated-Benford model for different choices of $X_{\mathrm{B}}$ in Equation (14) of the paper, when $n=200$. The exact test size is $\gamma=0.01$. Null quantiles of test statistics are estimated with $10^6$ replicates.}
	\label{tab:n200}

\begin{small}
\begin{center}
		\begin{tabular}{@{}cccccccccccc@{}}
	$\alpha$ & $Q_1$ & $Q_{12}$ & $Q_2$ & $KS_1$ & $\varLambda_{\widehat{N},n}$ & $KU_1$ & $KS_2$ & $KU_2$ & $Q_{\varDelta}$ & $G_{KS}$ & $G_{KU}$   \\ 
	\hline
	\multicolumn{12}{c}{$\mathrm{Lognormal}(\alpha,1)$}\\
    0.3   & 0.009 & 0.074 & 0.152 & 0.022 & 0.013 & 0.017 & 0.792 & 0.613 & 0.474 & 0.745 & 0.600 \\
    0.4   & 0.008 & 0.023 & 0.029 & 0.011 & 0.008 & 0.011 & 0.304 & 0.176 & 0.115 & 0.255 & 0.157 \\
    0.5   & 0.008 & 0.016 & 0.013 & 0.011 & 0.010 & 0.012 & 0.103 & 0.056 & 0.036 & 0.081 & 0.047 \\
    0.6   & 0.012 & 0.012 & 0.013 & 0.011 & 0.012 & 0.011 & 0.041 & 0.024 & 0.014 & 0.032 & 0.023 \\
	\hline		
	\multicolumn{12}{c}{$\mathrm{Weibull}(\alpha,1)$}\\
    2.2   & 0.011 & 0.018 & 0.018 & 0.013 & 0.010 & 0.012 & 0.144 & 0.076 & 0.041 & 0.112 & 0.062 \\
    2.6   & 0.011 & 0.028 & 0.036 & 0.012 & 0.011 & 0.009 & 0.346 & 0.209 & 0.133 & 0.289 & 0.191 \\
    3     & 0.008 & 0.046 & 0.070 & 0.014 & 0.012 & 0.014 & 0.589 & 0.420 & 0.264 & 0.527 & 0.390 \\
    3.4   & 0.011 & 0.084 & 0.146 & 0.020 & 0.014 & 0.021 & 0.796 & 0.652 & 0.426 & 0.744 & 0.612 \\
	\hline
	\multicolumn{12}{c}{$\mathrm{Uniform}(0, \alpha)$}\\
    5     & 0.012 & 0.018 & 0.067 & 0.011 & 0.025 & 0.023 & 0.124 & 0.062 & 0.227 & 0.198 & 0.180 \\
    20    & 0.009 & 0.022 & 0.056 & 0.010 & 0.023 & 0.019 & 0.121 & 0.060 & 0.222 & 0.191 & 0.168 \\
    40    & 0.011 & 0.021 & 0.068 & 0.011 & 0.022 & 0.020 & 0.119 & 0.058 & 0.231 & 0.202 & 0.183 \\
    60    & 0.010 & 0.018 & 0.059 & 0.011 & 0.026 & 0.023 & 0.120 & 0.061 & 0.222 & 0.193 & 0.177 \\
	\hline
	\multicolumn{12}{c}{$\mathrm{Generalized \; Benford}(\alpha)$}\\
    -1.0  & 0.009 & 0.034 & 0.048 & 0.012 & 0.009 & 0.014 & 0.423 & 0.228 & 0.202 & 0.371 & 0.231 \\
    1.0   & 0.009 & 0.018 & 0.067 & 0.012 & 0.024 & 0.024 & 0.117 & 0.060 & 0.228 & 0.199 & 0.178 \\
    2.0   & 0.009 & 0.034 & 0.142 & 0.013 & 0.027 & 0.030 & 0.301 & 0.163 & 0.440 & 0.415 & 0.378 \\
    3.0   & 0.008 & 0.044 & 0.230 & 0.018 & 0.036 & 0.041 & 0.472 & 0.278 & 0.602 & 0.590 & 0.546 \\
	\hline
\end{tabular}
\end{center}
\end{small}
\end{table*}


\begin{table*}[h!]
	\caption{Estimated power, based on 5000 simulations, under the manipulated-Benford model for different choices of $X_{\mathrm{B}}$ in Equation (14) of the paper, when $n=1000$. The exact test size is $\gamma=0.01$. Null quantiles of test statistics are estimated with $10^6$ replicates.}
	\label{tab:n1000}
\begin{small}
\begin{center}
		\begin{tabular}{@{}cccccccccccc@{}}
	$\alpha$ & $Q_1$ & $\chi^2_{\{12\}}$ & $Q_2$ & $KS$ & $\varLambda_{\widehat{N},n}$ & $KU$ & $KS_{fr}$ & $KU_{fr}$ & $Q_{\Delta}$ & $G_{KS_{fr}}$ & $G_{KU_{fr}}$   \\ 
	\hline
	\multicolumn{12}{c}{$\mathrm{Lognormal}(\alpha,1)$}\\
    0.3   & 0.010 & 0.950 & 0.976 & 0.078 & 0.023 & 0.082 & 1.000 & 1.000 & 1.000 & 1.000 & 1.000 \\
    0.4   & 0.010 & 0.354 & 0.357 & 0.023 & 0.012 & 0.022 & 0.983 & 0.936 & 0.740 & 0.971 & 0.911 \\
    0.5   & 0.011 & 0.089 & 0.053 & 0.019 & 0.014 & 0.019 & 0.632 & 0.432 & 0.181 & 0.564 & 0.368 \\
    0.6   & 0.009 & 0.046 & 0.015 & 0.013 & 0.014 & 0.016 & 0.232 & 0.127 & 0.031 & 0.184 & 0.093 \\
	\hline		
	\multicolumn{12}{c}{$\mathrm{Weibull}(\alpha,1)$}\\
    2.2   & 0.011 & 0.135 & 0.083 & 0.016 & 0.012 & 0.018 & 0.808 & 0.629 & 0.313 & 0.749 & 0.557 \\
    2.6   & 0.010 & 0.389 & 0.376 & 0.030 & 0.010 & 0.025 & 0.991 & 0.960 & 0.773 & 0.986 & 0.943 \\
    3.0   & 0.011 & 0.758 & 0.755 & 0.048 & 0.017 & 0.051 & 1.000 & 0.999 & 0.969 & 1.000 & 0.998 \\
    3.4   & 0.008 & 0.953 & 0.945 & 0.078 & 0.023 & 0.088 & 1.000 & 1.000 & 0.996 & 1.000 & 1.000 \\
	\hline
	\multicolumn{12}{c}{$\mathrm{Uniform}(0, \alpha)$}\\
    5     & 0.012 & 0.131 & 0.624 & 0.045 & 0.085 & 0.123 & 0.727 & 0.513 & 0.920 & 0.900 & 0.888 \\
    20    & 0.011 & 0.128 & 0.617 & 0.044 & 0.087 & 0.120 & 0.711 & 0.498 & 0.921 & 0.897 & 0.886 \\
    40    & 0.011 & 0.131 & 0.607 & 0.047 & 0.089 & 0.124 & 0.713 & 0.498 & 0.921 & 0.898 & 0.887 \\
    60    & 0.009 & 0.135 & 0.620 & 0.047 & 0.093 & 0.122 & 0.736 & 0.523 & 0.927 & 0.906 & 0.897 \\
	\hline
	\multicolumn{12}{c}{$\mathrm{Generalized \; Benford}(\alpha)$}\\
    -1.0  & 0.010 & 0.546 & 0.611 & 0.033 & 0.014 & 0.032 & 0.997 & 0.984 & 0.922 & 0.996 & 0.978 \\
    1.0   & 0.008 & 0.131 & 0.608 & 0.042 & 0.089 & 0.120 & 0.717 & 0.508 & 0.925 & 0.903 & 0.892 \\
    2.0   & 0.012 & 0.410 & 0.948 & 0.130 & 0.143 & 0.238 & 0.985 & 0.935 & 0.998 & 0.998 & 0.997 \\
    3.0   & 0.012 & 0.669 & 0.993 & 0.230 & 0.205 & 0.337 & 0.999 & 0.994 & 1.000 & 1.000 & 1.000 \\
	\hline
\end{tabular}
\end{center}
\end{small}
\end{table*}

\begin{table*}[h!]
	\caption{Proportion of rejections of $H_0$, based on 5000 simulations, under a truncated-Benford model with $k$ significant digits in $S(X)$, when $n=200,1000$. The exact test size is $\gamma=0.01$. Null quantiles are the same as in Tables \ref{tab:n200} and \ref{tab:n1000} of this Supplement.}
	\label{tab:trunc}

    \begin{center}
	\begin{tabular}{@{}cccccc@{}}
	$k$ & $KS_2$ & $KU_2$ & $Q_{\varDelta}$ & $G_{KS}$ & $G_{KU}$   \\ 
	\hline
	\multicolumn{6}{c}{$n=200$}\\	
    6     & 0.011 & 0.012 & 0.010 & 0.010 & 0.012 \\
    5     & 0.009 & 0.011 & 0.008 & 0.009 & 0.011 \\
    4     & 0.008 & 0.008 & 0.009 & 0.009 & 0.008 \\
    3     & 0.011 & 0.013 & 0.013 & 0.012 & 0.013 \\
    2     & 0.820 & 0.943 & 0.302 & 0.756 & 0.891 \\
	\hline
	\multicolumn{6}{c}{$n=1000$}\\	
    6     & 0.011 & 0.011 & 0.009 & 0.011 & 0.010 \\
    5     & 0.010 & 0.009 & 0.008 & 0.008 & 0.008 \\
    4     & 0.008 & 0.008 & 0.009 & 0.009 & 0.009 \\
    3     & 0.031 & 0.042 & 0.020 & 0.025 & 0.034 \\
    2     & 1.000 & 1.000 & 0.982 & 1.000 & 1.000 \\
	\hline
\end{tabular}
\end{center}

\end{table*}

\subsubsection{Effect of rounding}

Table \ref{tab:round} of this Supplement repeats the study of the effect of contamination in the last digits of $S(X)$ when rounding to the $k$th digit is performed instead of truncation. The sample sizes are $n=200$, $n=500$ and $n=1000$. 


\begin{table*}[h!]

\caption{Proportion of rejections of $H_0$, based on 5000 simulations, under a rounded-Benford model with $k$ significant digits in $S(X)$, when $n=200,500,1000$. The exact test size is $\gamma=0.01$. Null quantiles are the same as in Tables \ref{tab:n200} and \ref{tab:n1000} of this Supplement ($n=200,1000$) and in Table 3 of the paper ($n=500$).}
\label{tab:round}
        
\begin{center}
	\begin{tabular}{@{}cccccc@{}}
	$k$ & $KS_2$ & $KU_2$ & $Q_{\varDelta}$ & $G_{KS}$ & $G_{KU}$   \\ 
	\hline
	\multicolumn{6}{c}{$n=200$}\\	
    6     & 0.013 & 0.012 & 0.009 & 0.010 & 0.011 \\
    5     & 0.012 & 0.013 & 0.013 & 0.012 & 0.012 \\
    4     & 0.010 & 0.008 & 0.009 & 0.009 & 0.008 \\
    3     & 0.012 & 0.013 & 0.011 & 0.009 & 0.011 \\
    2     & 0.656 & 0.928 & 0.125 & 0.573 & 0.858 \\
	\hline
	\multicolumn{6}{c}{$n=500$}\\	
    6     & 0.011 & 0.011 & 0.009 & 0.010 & 0.010 \\
    5     & 0.009 & 0.010 & 0.008 & 0.008 & 0.009 \\
    4     & 0.012 & 0.010 & 0.011 & 0.012 & 0.010 \\
    3     & 0.016 & 0.024 & 0.011 & 0.014 & 0.018 \\
    2     & 1.000 & 1.000 & 0.410 & 1.000 & 1.000 \\
	\hline	
	\multicolumn{6}{c}{$n=1000$}\\	
    6     & 0.010 & 0.011 & 0.008 & 0.008 & 0.008 \\
    5     & 0.008 & 0.008 & 0.009 & 0.008 & 0.008 \\
    4     & 0.010 & 0.013 & 0.010 & 0.010 & 0.011 \\
    3     & 0.025 & 0.041 & 0.010 & 0.017 & 0.026 \\
    2     & 1.000 & 1.000 & 0.787 & 1.000 & 1.000 \\
	\hline
\end{tabular}
\end{center}

\end{table*}



\subsection{A refined simulation scheme in the case of truncation or rounding}

Following the notation of the paper, let $n$ be the sample size and $S(X_1), \ldots, S(X_n)$ be the significands of the sample values $X_1,\ldots,X_n$. For $k=1,\ldots,K$, we define $n_k$ to be the number of sample values with $k$ significant digits in their significand. It clearly holds that $\sum_{k=1}^K n_k \le n$, while the difference $n - \sum_{k=1}^K n_k$ yields the number of sample significands with more than $K$ significant digits. 
Our goal is to simulate $B$ samples of $n$ Benford random numbers, using properties \eqref{benf} and \eqref{significand} above, and to truncate (or round) them in order to match the observed pattern $n_1,\ldots,n_K$ in each simulated sample. This truncated-Benford (or rounded-Benford) scheme adapts the Benford hypothesis to practical situations where discretization may occur in the absence of substantial irregularities. The corresponding Monte Carlo estimates of the null quantiles of our test statistics will then be robust to such (possibly unimportant) violations of the Benford hypothesis due to discretization. The results displayed in 
Section 6
of the paper and in Tables \ref{tab:trunc} and \ref{tab:round} of this Supplement demonstrate that the effect of truncation (or rounding) on our tests is negligible unless $k$ is very small. We thus suggest to take $K=6$ for virtually all practical purposes.

Given a tail probability $\gamma$, the pseudocode of the modified Monte Carlo procedure for estimating the null quantile of a test statistic is provided 
as Algorithm \ref{alg1}, where we focus on truncation. The case of rounding can be dealt with in an analogous manner. Our algorithm requires sorting of the significands, so that $S_{(1)} \le \cdots \le S_{(n)}$ are the order statistics of $S(X_1), \ldots, S(X_n)$, as in the paper. Furthermore, we let $k_{(i)}$ denote the number of significant digits of $S_{(i)}$, so that for $k=1,\ldots,K$ and in the notation of the paper
\[
\sum_{i=1}^n \boldsymbol{1}_{\{k\}}(k_{(i)}) = n_k ,
\]
as required. Similarly, for each simulated sample 
\[
X_{b,1},\ldots,X_{b,n}
\]
and $b=1,\ldots,B$, we write $k_{b,(i)}$ for the number of significant digits of $S_{b,(i)}$, where $S_{b,(1)} \le \cdots \le S_{b,(n)}$ are the order statistics of $S(X_{b,1}), \ldots, S(X_{b,n})$.

\begin{algorithm*}[t!]
\vskip 1mm \noindent
\caption{Monte Carlo algorithm for quantile estimation under a truncated-Benford model\label{alg1}}
\begin{algorithmic}[1]
\State set $K$, $B$ and $\gamma$; read $k_{(1)},\ldots,k_{(n)}$
\For {$b \in \{1,\ldots,B\}$}
\For {$i \in \{1,\ldots,n\}$}
\State {simulate $U_{b,i}$ as independent Uniform random variables on $[0, 1)$}
\State {compute $S(X_{b,i})=10^{U_{b,i}}$}
\EndFor
\State {sort $S(X_{b,1}), \ldots, S(X_{b,n})$ to obtain $S_{b,(1)}\le \ldots\le S_{b,(n)}$}
\For {$i \in \{1,\ldots,n\}$}
\If {$k_{(i)}\le K$}
\State {truncate $S_{b,(i)}$ to make $k_{b,(i)}=k_{(i)}$; write $S_{b,(i)}^*$ for the truncated $S_{b,(i)}$}
\Else 
\State {$S_{b,(i)}^*=S_{b,(i)}$}
\EndIf
\EndFor
\State {compute the test statistic of interest $T^*_b=T(S_{b,(1)}^*,\ldots,S_{b,(n)}^*)$}
\EndFor
\State {compute the $(1-\gamma)$-quantile of $T^*_1,\ldots,T^*_B$}
\end{algorithmic}
\end{algorithm*}

\section{WebARIADNE revisited}


This supplement introduces and documents WebAriadne, an enhanced web application designed to make our proposed methodology accessible to practitioners and analysts. This part of the supplement, largely self-contained, offers additional context and motivation for the development of such tools. In particular, it highlights the growing demand for statistically sound and user-friendly platforms to support fraud detection in international trade and other domains. WebAriadne is specifically designed to bridge the gap between advanced statistical techniques and operational needs, reinforcing the practical relevance of the innovations presented in the main paper.

\subsection{Recalling WebARIADNE}
Originally introduced in a previous paper \citep{cer+al:19}, WebARIADNE is a web-based application designed to facilitate the detection of statistical anomalies and underlying structures in large-scale datasets, including customs data. What sets WebARIADNE apart is its dual approach to anomaly detection: it employs robust estimation methods to identify traditional statistical outliers -- values that significantly deviate from expected distributions -- while also leveraging Benford’s Law tests to uncover data fabrications that may not appear as conventional outliers. This combination enables the system to detect both unusual variations (legitimate or not) and systematic distortions characteristic of manipulated data. In addition to these advanced techniques, WebARIADNE integrates standard methodologies for data import, preprocessing, and user-defined descriptive statistics, facilitating structured data exploration. The system features a client-side JavaScript interface and a Java-based backend running on Apache Tomcat, serving as the central hub for client requests and statistical computations.  The architecture of the statistical engine integrates multiple  environments: SAS, MATLAB, R and now also Python. This integration allows teams to leverage the strengths of each language, such as SAS's enterprise-grade scalability and Python's AI capabilities, to optimally match development tools with application requirements and improved computational efficiency.
This section illustrates the enhancements introduced in recent years to align WebARIADNE with the EU’s push for greater use of digital tools in customs enforcement. 

%

\subsection{Need for new analytic capabilities in the  evolving EU Customs Union}

After five years from the European Commission President plan ``to take the Customs union to the next level’’, this vision of both political and technical significance has become even more imperative in light of recent geopolitical transformations, particularly the Russian invasion of Ukraine and the subsequent sanctions regimes, as well as the resurgence of protectionist trade measures, such as tariffs on goods. Indeed, a potential consequence of protectionist trade measures is that they create additional dishonest incentives for trade misreporting: in addition to claiming a value below its actual worth, which we have extensively discussed in this contribution and other documents \citep{per+al:20,PCTCA:2020}, other prevalent tactics include the incorrect classification of product goods (for instance, labeling steel as machinery parts to circumvent anti-dumping duties) and transshipment through third countries, falsely declaring the origin of goods to evade country-specific restrictions. Furthermore, even official trade statistics could be possibly susceptible to manipulation by countries seeking to downplay economic distress or conceal sanctioned (or even illicit) trade flows \citep{fis+al:09}. A recent high-level policy report \cite{wisepersons2022} wrote by independent experts identifies the key challenges confronting the EU Customs Union and proposes a comprehensive reform agenda aimed at modernizing its operations. Their recommendations place a strong emphasis on the necessity of effectively exploiting data and digital tools for enhanced customs risk management.

\subsection{Key innovations and enhancements}
Against this backdrop, WebARIADNE and its counterpart THESEUS form a complementary system for strengthening the EU’s customs capabilities. WebARIADNE in support to the detection of statistical anomalies, and THESEUS to the dissemination of relevant signals to customs officers, with the latter explicitly recognized by the European Court of Auditors \citep{ECA2017} as an effective tool for enhancing customs risk management.
 Over the past years, the growing complexity of international trade because of the above context and the rise of e-commerce, have necessitated significant advancements in such tools.
The updated WebARIADNE system integrates more sophisticated statistical techniques, with enhanced anomaly detection, and improved data visualization capabilities.  

\begin{figure}[t!]
\begin{minipage}{\textwidth}
\fbox{\includegraphics[width=0.475\textwidth]{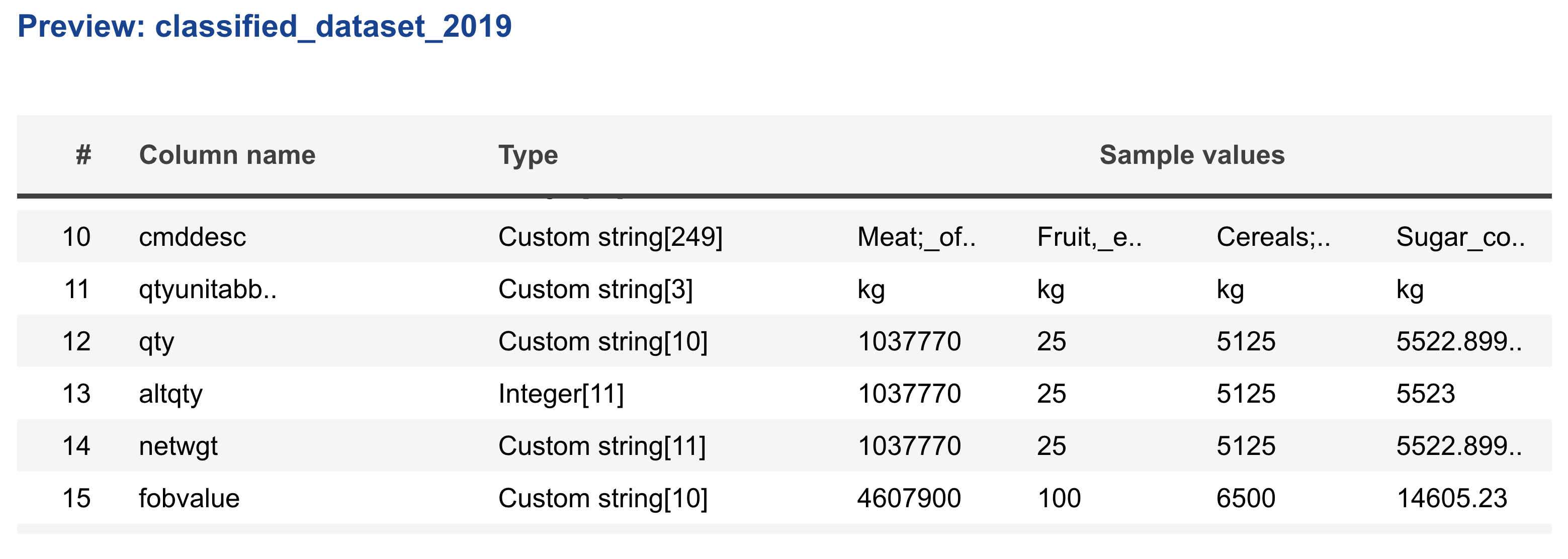}}
\hspace{22mm}
\fbox{\includegraphics[width=0.305\textwidth]{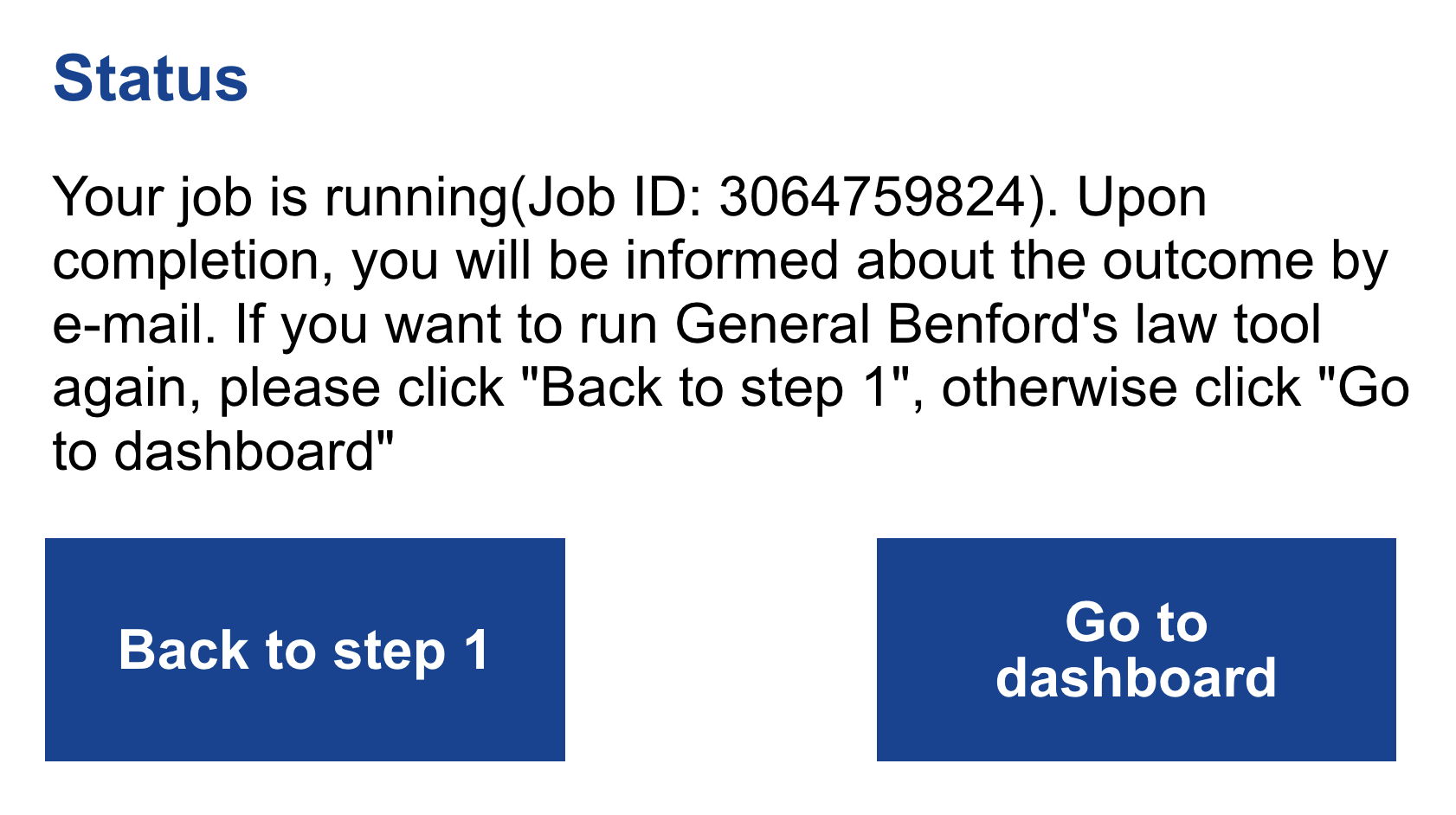}}
\end{minipage}
\vskip 5mm
\begin{minipage}{\textwidth}
\fbox{\includegraphics[width=0.475\textwidth]{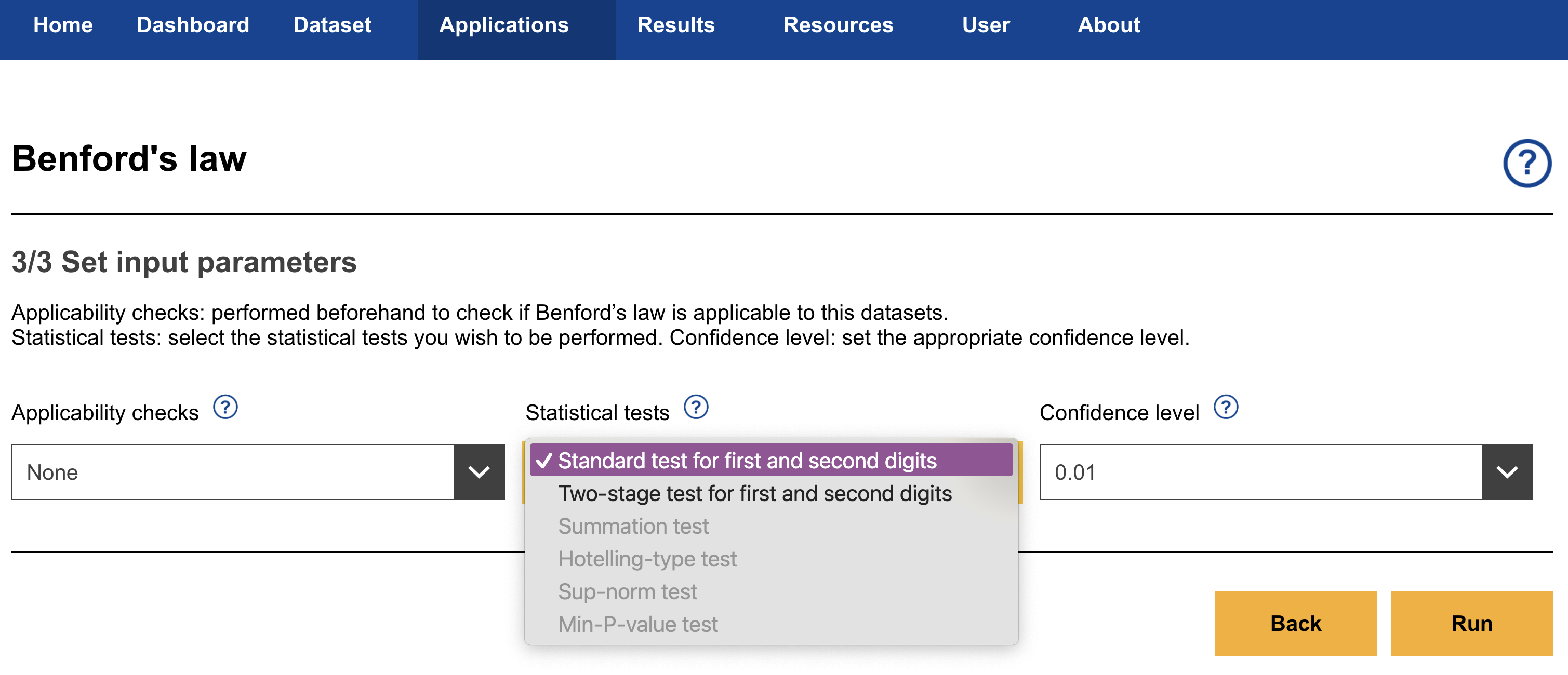}}
\hfill
\fbox{\includegraphics[width=0.465\textwidth]{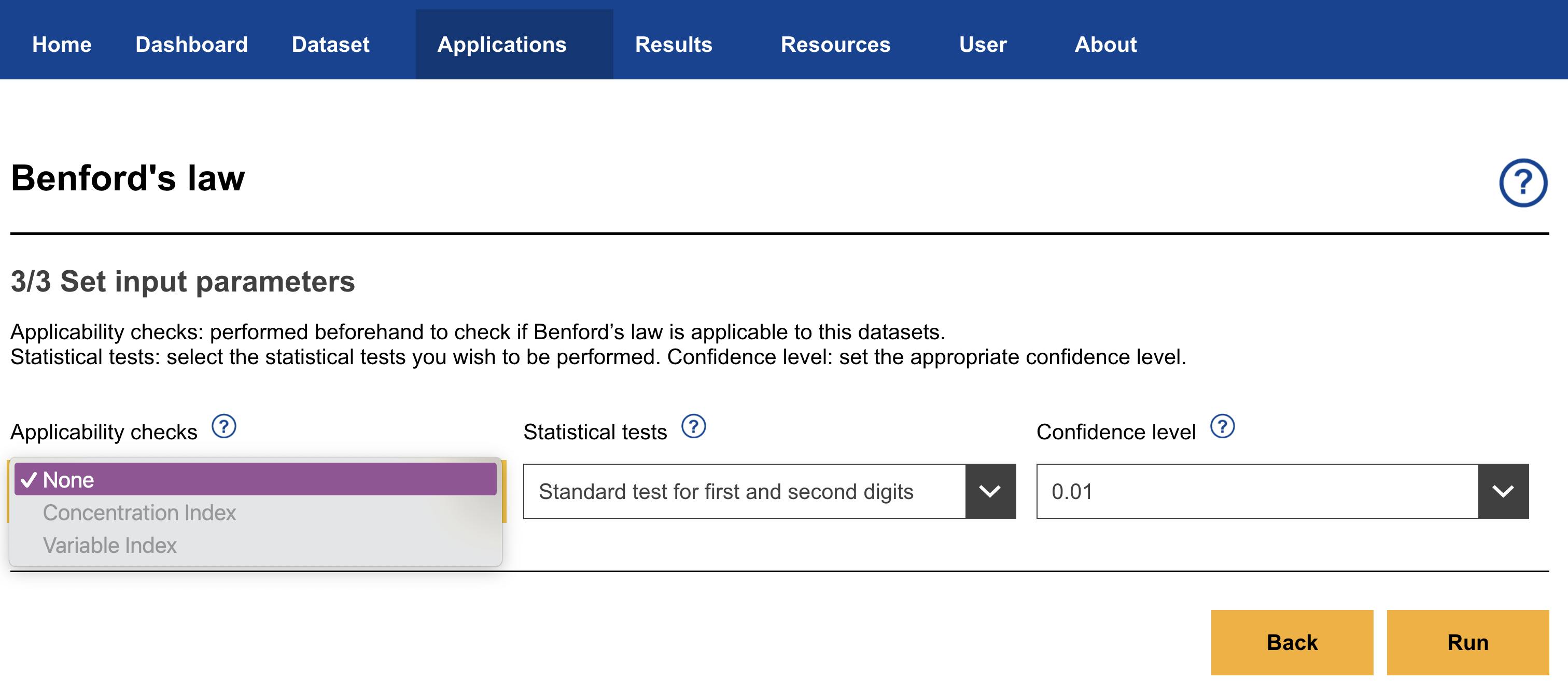}}
\caption{\label{WA_features:fig} Some WebARIADNE features. Top panels: preview of data types found in an imported dataset (left) and dialogue box of the asynchronous execution modality for large datasets (right). Bottom panels: selection of Benford's law input parameters (type of statistical test, confidence level and applicability filters).} 
\end{minipage}
\vskip 15mm
\begin{minipage}{\textwidth}
\fbox{\includegraphics[width=0.473\textwidth]{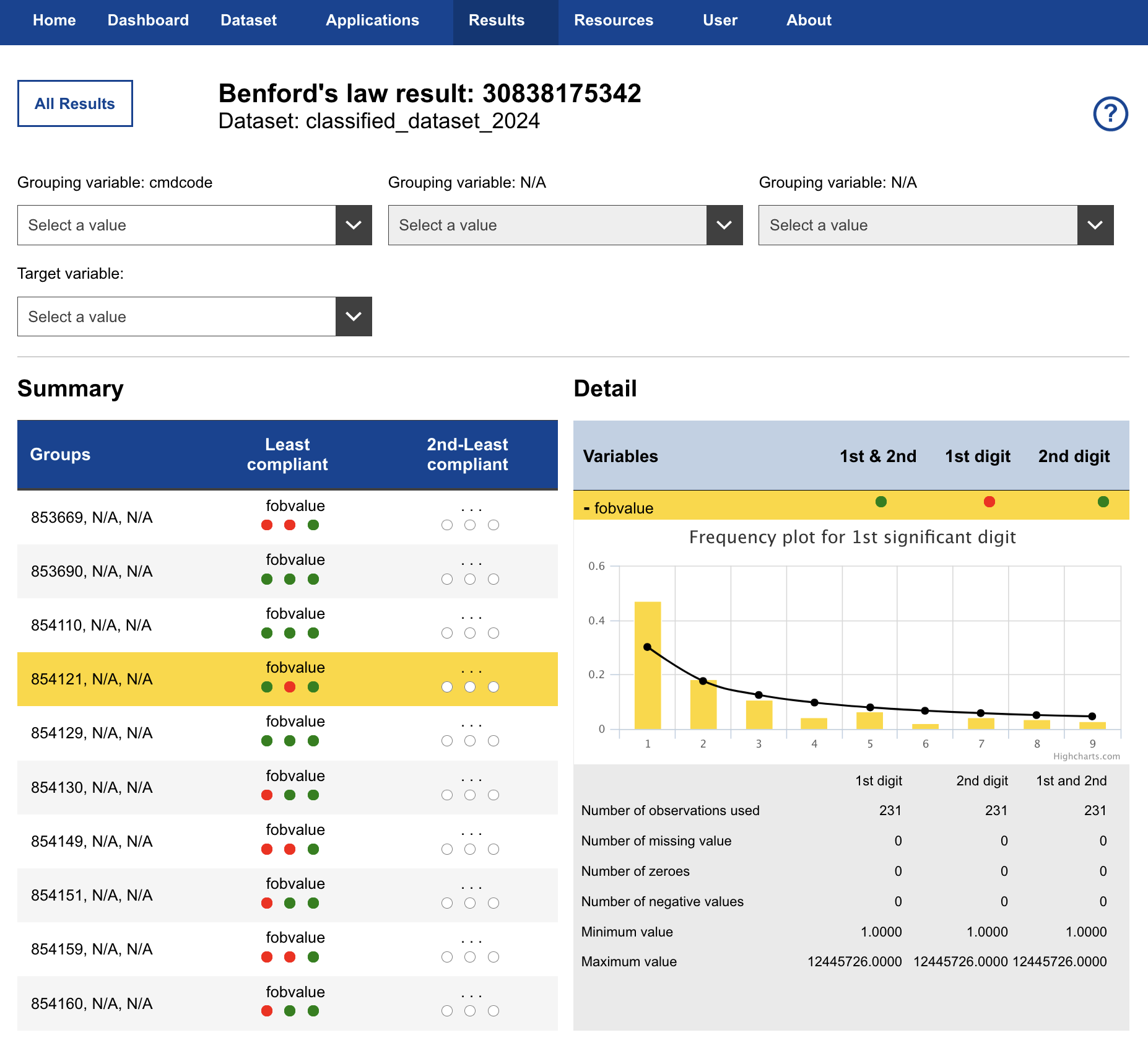}}
\hfill
\fbox{\includegraphics[width=0.47\textwidth]{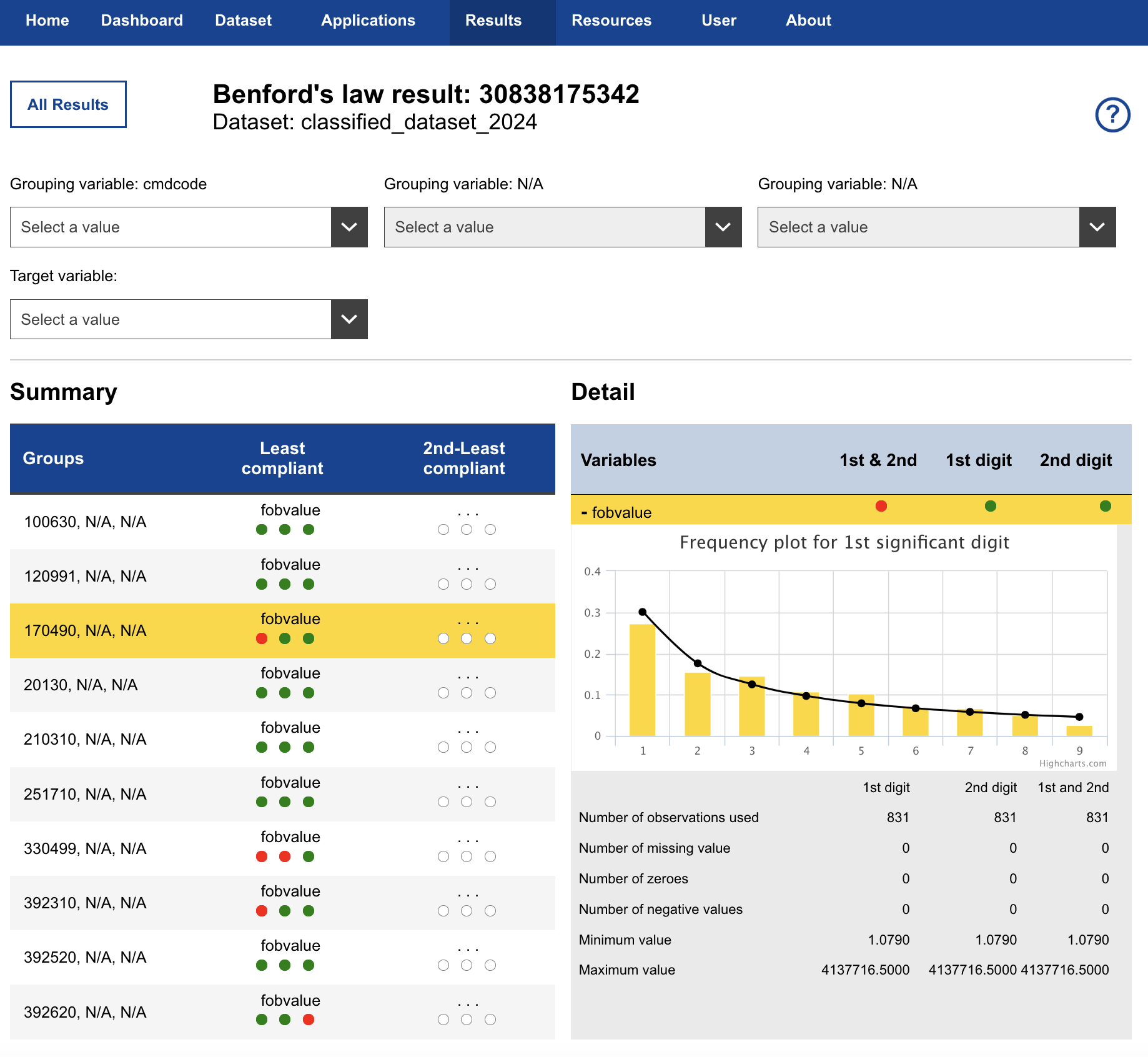}}
\caption{\label{WA_BenfPlots:fig}Benford's results on a collection of datasets. Left panel: test rejection on the first digit. Right panel: test rejection on the first two digits, but not on the individual ones. } 
\end{minipage}
\end{figure}

One notable enhancement is the expansion of trade data sources, enabling cross-validation of both micro-level customs declarations and macro-level trade statistics. WebARIADNE can access these sources from an internal database, but users can also upload any dataset of interest, partition the data into samples by defining specific grouping variables, and execute a detection tool on all samples simultaneously. In this process, users are supported by dialogue boxes to inspect the data and analyze them asynchronously for large data collections, with email notification capabilities (top panels of Figure \ref{WA_features:fig}).

\begin{figure}[h]
\fbox{\begin{minipage}{0.48\textwidth}
\centering
\includegraphics[width=0.97\textwidth]{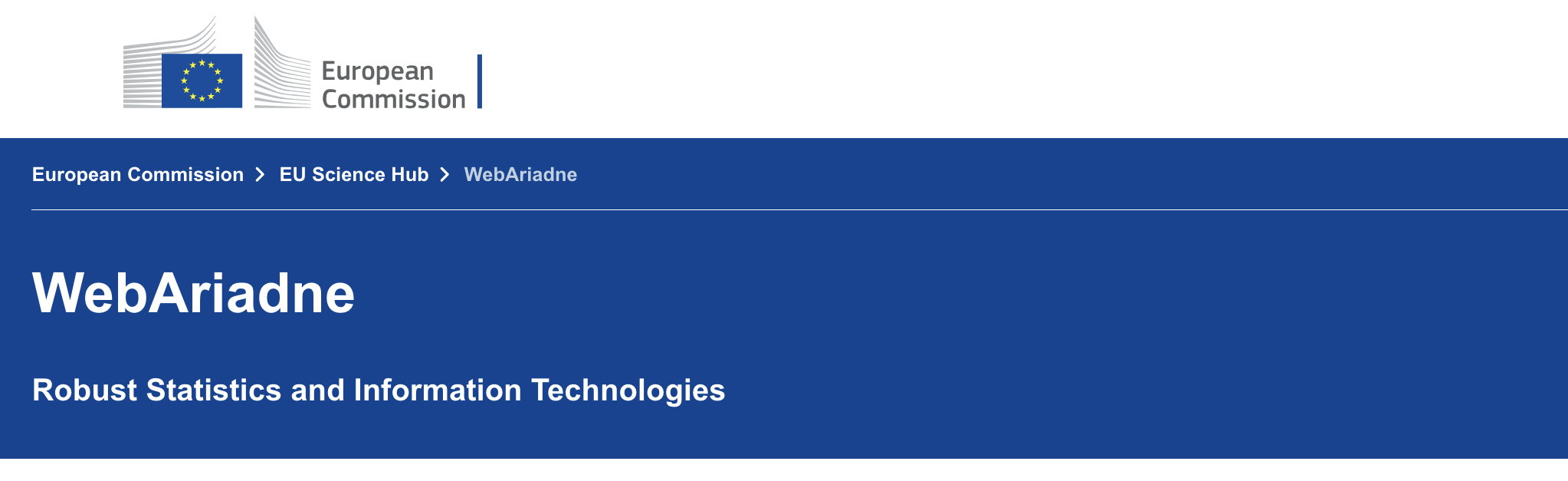}\\
\includegraphics[width=\textwidth]{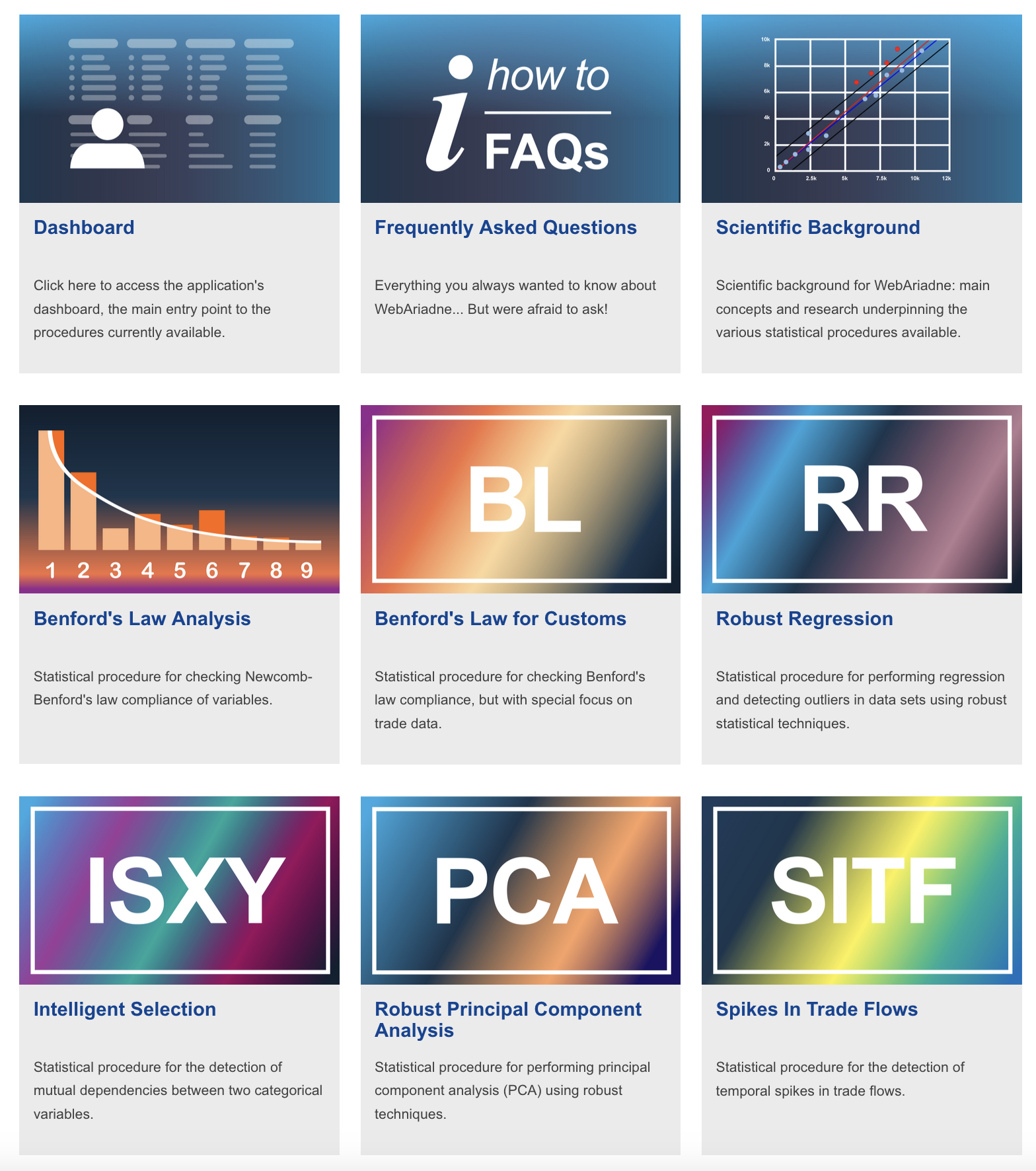}
\end{minipage} 
}
\hfill
\begin{minipage}{0.48\textwidth}
\centering
\fbox{\includegraphics[width=0.95\textwidth]{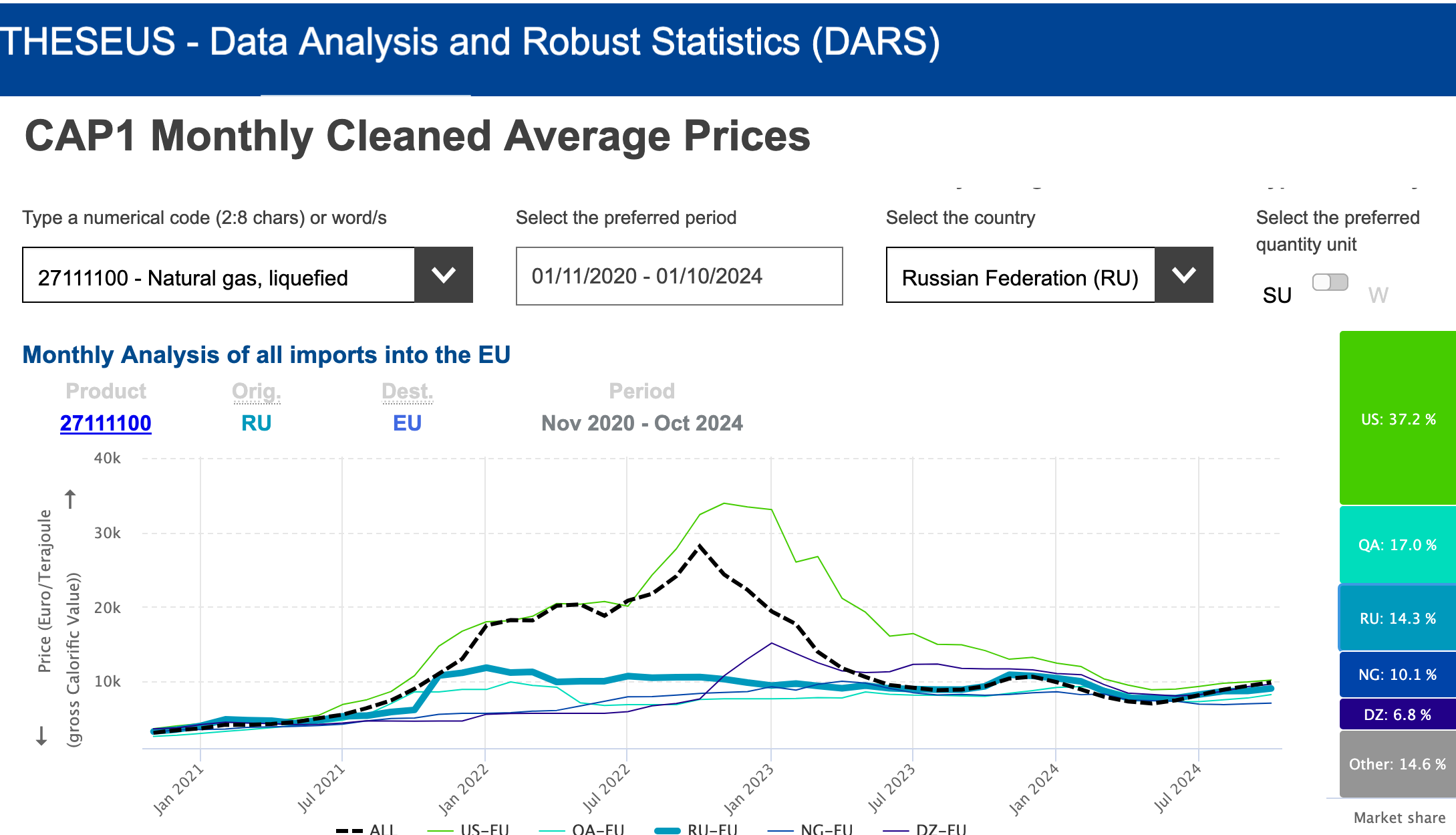}}\\[18mm]
\fbox{\includegraphics[width=0.95\textwidth]{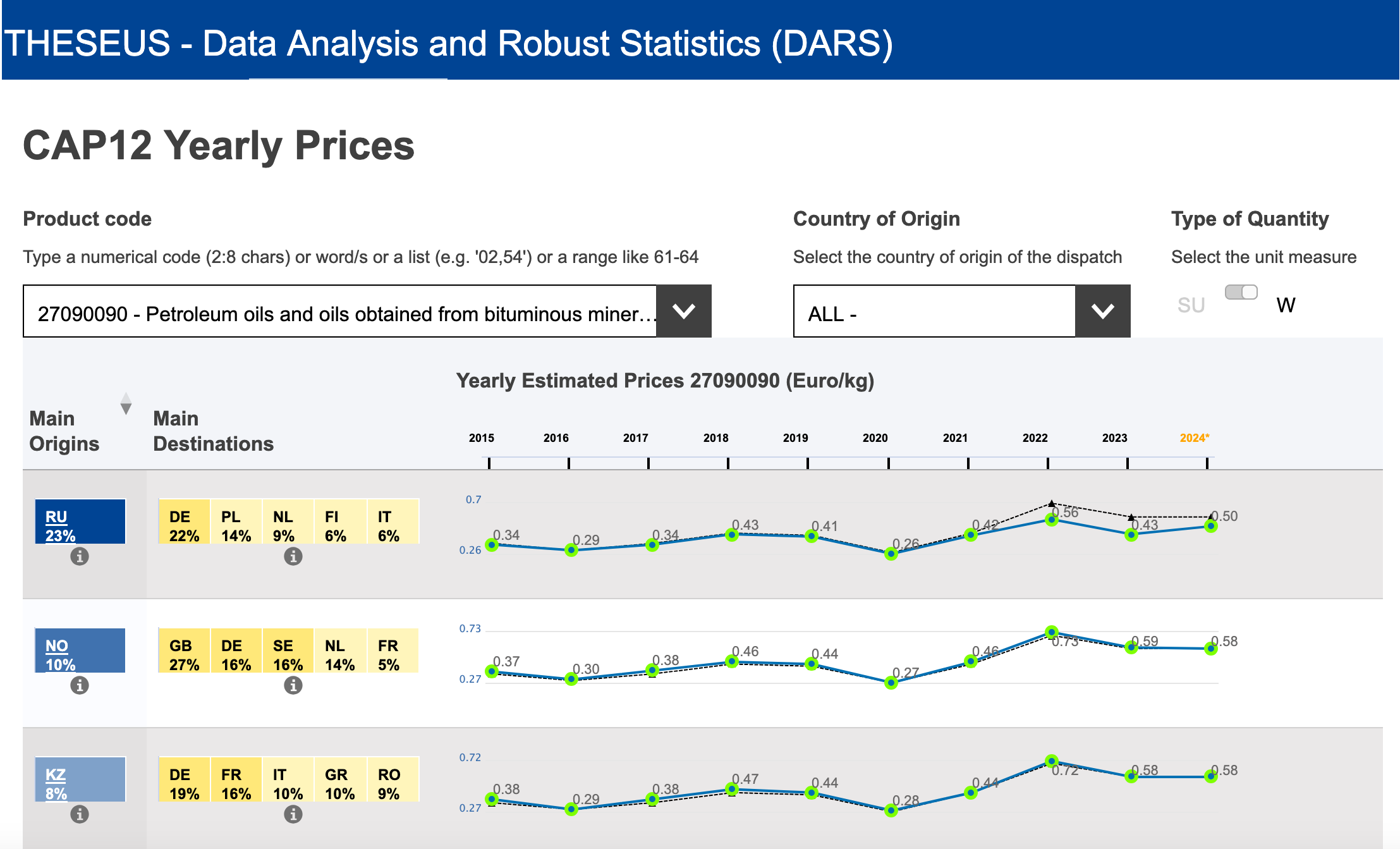}}
%
%
\end{minipage}
%
\caption{\label{WA_TH_landigpages:fig}Left panel: the landing page of WebARIADNE with the pointers to its main statistical applications. Right panel: two examples of THESEUS dashboards for the dissemination of trade prices estimated by applying tools available in WebARIADNE on internal and public data. Both systems are accessible by authorized users only.}
\end{figure}

A second significant enhancement is the integration of novel, robust statistical models that now complement Benford's law-based techniques for enhancing the overall fraud detection rates. Some of these models are highlighted on the landing page of WebARIADNE, as depicted in Figure \ref{WA_TH_landigpages:fig}. The right panels of the same figure showcase dissemination dashboards of THESEUS. These two platforms are part of a coherent digital platform ecosystem. For instance, the price estimates shown in the two THESEUS examples utilize the estimation method provided by the `Robust Regression' module in WebARIADNE.

Again on the Benford's law module, there is the possibility of selecting different testing methods. These methods include those introduced in \cite{bar+al:16b}, \cite{cer+al:19} and shortly the one presented in this article, which is currently being implemented. The bottom panels of Figure \ref{WA_features:fig} demonstrate how the user can select the appropriate test, as well as the desired confidence level and the application of tools for screening the data for potential issues that could prevent the application of Benford's law, such as the concentration of repeated data values.
The inspection of the results of the application of Benford's module on large collections of trade flows samples is simplified by intuitive dashboards, exemplified by Figure \ref{WA_BenfPlots:fig}, reporting the detailed statistics obtained in each test as well as a graphical visualization of the typical Benford's histogram on the first significant digit. Note that the list of results obtained on the full dataset collection is synthetically previewed in a single ``Summary'' tab, giving an overall idea of the problems detected in the collection.
The results displayed in the left panel of Figure \ref{WA_BenfPlots:fig} -- if the conditions that anticipate validity of Benford's holds in the case under scrutiny -- can provide an instance of ``naive'' data fabrication in the first digit of traded values, highlighted by the classical first-digit Pearson test. In the right panel the first-digit test does not reject the Benford hypothesis, which is instead rejected by the two-digit test \eqref{Q12}. It thus provides motivation for the work in this paper, showing that more subtle data manipulations can occur in the digits following the first one.

In conclusion, we believe that the practical applicability of the novel contribution to Benford's testing presented in this paper will be enhanced by its integration into WebARIADNE, facilitated by its enhanced usability and more user-friendly analytics, expanded data capabilities to process efficiently larger datasets, novel formats, and more complex structures.


\end{document}